\documentclass{elsart}
\usepackage{makeidx}
\usepackage{longtable}

 \usepackage{epsfig}
\usepackage{srcltx}
\usepackage{cite,slashed,axodraw}
\usepackage{amssymb}
\makeindex

\begin{document}
\newcounter{im}
\setcounter{im}{0}
\newcommand{\exampleSp}{\stepcounter{im}\includegraphics[scale=0.9]{SpinorExamples_\arabic{im}.eps}}
\newcommand{\myindex}[1]{\label{com:#1}\index{{\tt #1} & pageref{com:#1}}}
\renewcommand{\topfraction}{1.0}
\renewcommand{\bottomfraction}{1.0}
\renewcommand{\textfraction}{0.0}
\newcommand{\nc}{\newcommand}
\nc{\eqn}[1]{Eq.~\ref{eq:#1}}
\nc{\be}{\begin{equation}}
\nc{\ee}{\end{equation}}
\nc{\ba}{\begin{array}}
\nc{\ea}{\end{array}}
\nc{\bea}{\begin{eqnarray}}
\nc{\eea}{\end{eqnarray}}

\nc{\semi}{;\ }
\nc{\spa}[3]{\left\langle#1\,#3\right\rangle}
\nc{\spb}[3]{\left[#1\,#3\right]}
\nc{\Tr}{\mathop{\rm Tr}\nolimits}
\nc{\hf}{\textstyle{1\over2}}
\nc{\si}{\sigma}
\nc{\ns}{n_{\mskip-2mu s}}
\nc{\nf}{n_{\mskip-2mu f}}
\nc{\ib}{{\bar\imath}}
\nc{\jb}{{\bar\jmath}}
\nc{\ksl}{\not{\hbox{\kern-2.3pt $k$}}}
\nc{\Slash}[1]{\slash\hskip -0.17 cm #1}
\renewcommand{\d}{\mathrm{d}}
\newcommand{\arctanh}{\mathrm{arctanh}}
\newcommand{\arccot}{\mathrm{arccot}}
\newcommand{\arccsc}{\mathrm{arccsc}}
\newcommand{\sgn}{\mathrm{sgn}}
\newcommand{\prim}[1]{{#1^{\prime}}}
\newcommand{\tr}{\mathrm{Tr}}
\newcommand{\naeher}{\!\!\!}
\newcommand{\vect}[1]{\vec #1}
\newcommand{\ddd}[1]{\d\vect{#1} }
\def\exboxlength{1.1\textwidth}

%% Pierpaolo Def's %%
\def\spa#1.#2{\langle#1\,#2\rangle}
\def\spb#1.#2{[#1\,#2]}

\def\spab#1.#2.#3{\langle\mskip-1mu{#1}
                  | #2 | {#3}]}

\def\spba#1.#2.#3{[\mskip-1mu{#1}
                  | #2 | {#3}\rangle}

\def\spbb#1.#2.#3.#4{[\mskip-1mu{#1}
                     | {#2} \ {#3} | {#4}]}

\def\spaa#1.#2.#3.#4{\langle\mskip-1mu{#1}
                     | {#2} \ {#3} | {#4}\rangle}

\def\dea{\langle \lambda \ d \lambda \rangle}
\def\deb{[\lambda \ d \lambda]}

\def\dedea{\langle d \lambda \ \partial_{|\lambda\ra} \rangle}
\def\dedeb{[d \lambda \ \partial_{|\lambda]} ]}
%%%%%%%%%%%%%%%%%%%%%%%%

\begin{frontmatter}


 \title{\vspace{-2.5cm}\hfill {\small\rm SLAC-PUB-12867, ZU-TH 25/07}\vspace{1cm}\\S@M, a Mathematica Implementation of the Spinor-Helicity Formalism}
 \author[uni,slac]{D. Ma\^{\i}tre}
%\thanksref{label1}
 \author[uni]{, P. Mastrolia}
%\thanksref{label2}}
% \ead{maitreda@slac.stanford.edu}
% \thanks[label1]{Stanford Linear Accelerator Center, University of Z\"urich}
% \thanks[label2]{University of Z\"urich}
% \corauth[cor1]{maitreda@slac.stanford.edu}
 \address[uni]{Institut f\"ur Theoretische Physik\\
 University of Z\"urich\\Winterthurerstrasse 190, CH-8057 Z\"urich }
 \address[slac]{Stanford Linear Accelerator Center\\2575 Sand Hill Road\\ Menlo Park, CA 94025}
 %\thanks[label3]{adress}

%\title{S@m, a Mathematica Implementation of the Spinor Formalism}


%\author{}

%\address{}

\begin{abstract}
%In this paper we present the Mathematica package S@M
%(Spinors@Mathematica) which allows to exploit the benefits of
%complex-spinor algebra within the context of on-shell technology along
%with the multi-purpose features of Mathematica. The package defines the
%spinor objects with their basic properties along with functions to
%manipulate them and offers the possibility of  evaluating
%them numerically at every computational step.
In this paper we present the package S@M (Spinors@Mathematica) which implements the spinor-helicity formalism in Mathematica. The package allows the use of complex-spinor algebra along with the multi-purpose features of Mathematica.
The package defines the spinor objects with their basic properties along with functions to manipulate them. It also offers the possibility of evaluating
the spinorial objects numerically at every computational step. The package is therefore well suited to be used in the context of on-shell technology, in particular for the evaluation of scattering amplitudes
at tree- and loop-level.\end{abstract}

\end{frontmatter}
{\bf PROGRAM SUMMARY/NEW VERSION PROGRAM SUMMARY}
  %Delete as appropriate.

\begin{small}
\noindent
{\em Manuscript Title:} S@M, a Mathematica Implementation of the Spinor-Helicity Formalism                                       \\
{\em Authors:} D. Ma\^{\i}tre, P. Mastrolia                                                \\
{\em Program Title:S@M}                                          \\
%{\em Journal Reference:}                                      \\
  %Leave blank, supplied by Elsevier.
%{\em Catalogue identifier:}                                   \\
  %Leave blank, supplied by Elsevier.
%{\em Licensing provisions:} none                                   \\
  %enter "none" if CPC non-profit use license is sufficient.
{\em Programming language: } Mathematica                                   \\
%{\em Computer:}                                               \\
  %Computer(s) for which program has been designed.
%{\em Operating system:}                                       \\
  %Operating system(s) for which program has been designed.
%{\em RAM:} bytes                                              \\
  %RAM in bytes required to execute program with typical data.
%{\em Number of processors used:}                              \\
  %If more than one processor.
%{\em Supplementary material:}                                 \\
  % Fill in if necessary, otherwise leave out.
  % Please give some freely chosen keywords that we can use in a
  % cumulative keyword index.
  % see http://www.aip.org/pacs/pacs.html 
{\em Classification:} 4.4 Feynman diagrams, 5 Computer Algebra, 11.1  General, High Energy Physics and Computing                                        \\
  %Classify using CPC Program Library Subject Index, see (
  % http://cpc.cs.qub.ac.uk/subjectIndex/SUBJECT_index.html)
  %e.g. 4.4 Feynman diagrams, 5 Computer Algebra.
%{\em External routines/libraries:}                                      \\
  % Fill in if necessary, otherwise leave out.
%{\em Subprograms used:}                                       \\
  %Fill in if necessary, otherwise leave out.
%{\em Catalogue identifier of previous version:}               \\
  %Only required for a New Version summary, otherwise leave out.
%{\em Journal reference of previous version:}                  \\
  %Only required for a New Version summary, otherwise leave out.
%{\em Does the new version supersede the previous version?:}    \\
  %Only required for a New Version summary, otherwise leave out.

{\em Nature of problem:} Implementation of the spinor-helicity formalism\\
  %Describe the nature of the problem here.
% 
%   \\
{\em Solution method:} Mathematica implementation\\
  %Describe the method solution here.
%Mathematica implementation  
%   \\
%{\em Reasons for the new version:}\\
  %Only required for a New Version summary, otherwise leave out.
%   \\
%{\em Summary of revisions:}\\
  %Only required for a New Version summary, otherwise leave out.
%   \\
%{\em Restrictions:}\\
  %Describe any restrictions on the complexity of the problem here.
%   \\
%{\em Unusual features:}\\
  %Describe any unusual features of the program/problem here.
%   \\
%{\em Additional comments:}\\
  %Provide any additional comments here.
%   \\
%{\em Running time:}\\
  %Give an indication of the typical running time here.
%   \\
%{\em References:}
%\begin{refnummer}
%\item Reference 1         % This is the reference list of the Program Summary
%\item Reference 2         % Type references in text as [1], [2], etc.
%\item Reference 3         % This list is different from the bibliography, which
                          % you can use in the Long Write-Up.
%\end{refnummer}

\end{small}
\newpage
\hspace{1pc}
{\bf LONG WRITE-UP}

% main text
\section{Introduction}
Theoretical understanding of background processes is essential to
single out interesting signals in the rich landscape of events which
will take place at the forthcoming CERN experiment, the Large Hadron
Collider (LHC).
% Feynman diagrams {\it describe} the perturbative
%development of the interactions among colliding particles, and
%the evaluation of scattering amplitudes usually 
%requires the systematic computation 
%of a large number of diagrams. 
%They represent purely rational functions 
%of the external invariants, in case of {\it leading order} (LO) processes, and
%rather complicated integrals, resulting in the product of rational 
%coefficients times transcendental functions, when contributions beyond 
%the LO are taken into account.\\
%
Many methods are available for computing Standard Model backgrounds at
the leading order (LO) in perturbation theory 
\cite{MADGRAPH,CompHEP,AMEGIC,ALPGEN,HELAC}, based either on
automatic summation of tree-level 
Feynman diagram, or  
off-shell recursive 
algorithms for currents~\cite{BGRecursion,VECBOS,LaterRecursive}.
But quantitative estimates for most processes require a
calculation with next-to-leading order (NLO) accuracy - 
see for instance~\cite{LesHouches2005}.  
NLO calculations require knowledge of both virtual and real-radiation
corrections to the LO process. While the real-radiation corrections 
can be computed using tree-level techniques,
the bottleneck for the availability of results with NLO level accuracy~\cite{ttH,NLOThreeJet,MCFM,HbbWbb,EGZHjj,NLOttjet,LMPTriboson,Lazopoulos:2007bv,Dittmaier:2007th,Campbell:2007ev}
is %represented by 
the non-trivial evaluation of one-loop virtual corrections.
New approaches tackling the evaluation of one-loop multi-parton amplitudes
have recently been under intense 
development~\cite{Davydychev,Passarino,GRACEQCD,
DDReduction,
GieleGloverNumerical,AguilaPittau,EGZH4p,BGHPS,EGZSemiNum, 
EGZ6g, XYZ, BinothRat,OPP,Ellis:2007br,NSSubtract,LMPTriboson,Lazopoulos:2007bv,
Neq4Oneloop,Neq1Oneloop,DDimU, UnitarityMachinery,
NeqOneNMHVSixPt, BCFGeneralized, BBDPSQCD, BBCF, OnShellOneLoop, Qpap,
Bootstrap, Genhel,LoopMHV, BFM,ABFKM,MastroliaTriple,
BFMassive}.

The spinor-helicity formalism~\cite{SpinorHelicity}
for scattering amplitudes
has %been 
proven an invaluable tool in perturbative computation since
its development in the 1980's, 
being responsible for the %existence 
discovery of compact representations 
of tree and loop amplitudes.  
Instead of Lorentz inner products of momenta, it relies on the
more fundamental spinor products. These neatly capture the analytic properties of on-shell scattering amplitudes, like the factorization
behavior on multi-particle-channels.  
%
%The recent boost in the progress of evaluating on-shell scattering amplitudes is due to the exploitation of qualitative information on their analytic properties,such as factorization and unitarity, which have been quantitatively turned into tools for computing them.
The recent boost in the progress of evaluating on-shell scattering amplitudes is due to turning qualitative information on their analytic properties into quantitative tools for computing them.

On-shell methods~\cite{Bern:2007dw} restrict the propagating states 
to the physical ones and the spinor-helicity formalism is therefore well suited
to avoid the (intermediate) treatment of unphysical degrees of freedom 
whose effects disappear from final results. Moreover,
on-shell methods are tailored for the {\it parallel} treatment
of sets of diagrams which share a common kinematic structure, 
such as multi-particle
poles at tree-level and branch-cuts at loop-level~\cite{BernChalmers,SplitUnitarity}.
They are therefore suitable for extracting analytic 
information from simpler amplitudes in a recursive/iterative fashion,
since the singularities of scattering amplitude are determined 
by lower-point amplitudes in the case of poles and by lower-loop
ones in the case of cuts~\cite{TreeReview,LanceTASI,OneLoopReview}. \\
On-shell methods were originally used in~\cite{Zqqgg} and 
in the more recent systematized implementations for
the completion of all
six-gluon helicity amplitudes~\cite{Neq4Oneloop,Neq1Oneloop,
NeqOneNMHVSixPt,BBDPSQCD,BBCF,BFM,Genhel,LoopMHV,XYZ}
and the calculation of the six-photon 
amplitudes~\cite{BinothRat,BinothPhotons} in agreement with 
the numerical results of~\cite{EGZ6g,Ellis:2007br} 
and ~\cite{NS6ph,OPPPhotons} respectively.

The singularity information can be extracted
by defining amplitudes for suitable complex, yet
on-shell, values of the external momenta - an idea that
stemmed from Witten's development of twistor string 
theory~\cite{WittenTopologicalString,Penrose,CSLectures,TwistorReview}.
Generating complex momenta by modifying
spinor variables, considered as fundamental objects, 
leads to new ways to exploit the kinematic properties 
of helicity amplitudes.
The new-born complex momenta have the property
of preserving overall momentum conservation and on-shell nature. 
Complex kinematics allow the exploration 
of singularities of on-shell amplitudes and the use of
factorization information to reconstruct tree amplitudes recursively
from their poles.
The application of factorization in the on-shell method is realized through 
gluing lower-point tree amplitudes to form higher points ones
linked by on-shell yet complex propagating particles.
The construction of tree
amplitudes via on-shell recursion essentially amounts to
a reversal of the {\it collinear limit}. This is made possible by complexifying momenta in their spinorial representation, and results in 
a quadratic recursion, the BCFW recurrence relation~\cite{BCFW}, which
works for massive particles~\cite{BGKS,MassiveRecursion,SW,Hall:2007mz} as well.

Complex kinematics are useful for the fulfillment of generalized 
unitarity conditions as well. At one
loop, generalized unitarity corresponds
to requiring {\it more} than two internal particles to be on-shell,
such constraints cannot be realized in general 
with real Minkowski momenta. \\
The application of unitarity as an on-shell method of calculation
is based on two principles:
{\it i)} sewing tree amplitudes together to form one-loop amplitudes;
{\it ii)} decomposing loop-amplitudes in terms of a basis
of scalar loop-integrals~\cite{BDKBasicIntegrals,OtherIntegrals}.
Matching the generalized cuts of the amplitude with the cuts of 
basic integrals provides an efficient way to 
 extract the {\it rational} coefficients from the decomposition.
The unitarity method~\cite{Neq4Oneloop,Neq1Oneloop} provides a 
technique for producing functions with the correct branch cuts in
all channels~\cite{OldUnitarity}, as determined 
by products of tree amplitudes.

The use of four-dimensional
states and momenta in the cuts enable the construction of the 
(poly)logarithmic terms in the amplitudes,
but generically drops rational
terms, which have to be recovered independently. \\
More recent improvements to the unitarity method~\cite{BCFGeneralized}  
use complex momenta within generalized
unitarity, allowing for a simple and purely algebraic determination of
box integral coefficients from quadruple-cuts.  
Using double- and triple-unitarity cuts have led to very efficient techniques for extracting triangle and bubble integral coefficients analytically
~\cite{BBCF,BFM,MastroliaTriple,FordeTriBub}.
In particular in~\cite{BBCF,BFM,MastroliaTriple} the phase-space 
integration has been reduced to the extraction of residues 
in spinor variables and, at the occurrence, 
to trivial Feynman-parametric integration.
This approach has been used to compute analytically some contributions to 
the six-gluon amplitude~\cite{BBCF,BFM},
and the calculation of the complete six-photon amplitudes~\cite{BinothPhotons},
whose cut-constructibility was shown in~\cite{BinothRat} .
Other approaches have combined the knowledge
of the generic structure of loop-{\it integrand}~\cite{Pittau,AguilaPittau} 
with the simplification induced by cut-constraints, ending up with
a unitarity-motivated loop-integral decomposition
~\cite{OPP,OPPPhotons,FordeTriBub,Ellis:2007br}.
%%%%%%%%%%%%%%%%%%%%

The four-dimensional version of the unitarity method leaves the pure rational-function terms in the amplitudes undetermined.  
New approaches to computing rational terms use an optimized organization
of Feynman diagrams, by focusing the standard tensor reduction to 
tensor integrals which could generate the rational
terms~\cite{XYZ,BinothRat}. \\
A recent investigation \cite{BSTZ} on the source of rational terms
has been exploring the idea 
of their generation via
%of generating them by using 
a set of Lorentz-violating counterterms.  \\
Alternatively, these rational
functions can be characterized by their kinematic poles.  An efficient
means for constructing these terms from
their poles and residues is based on BCFW-like recursion 
relations~\cite{OnShellOneLoop,Qpap,Bootstrap,Genhel,LoopMHV}.\\
The rational parts of amplitudes can be also detected with 
$D$-dimensional unitarity 
cuts~\cite{VanNeervenUnitarity,DDimU,OneLoopReview,BMST,ABFKM},
and in~\cite{ABFKM,BFMassive} the benefits of four-dimensional 
spinorial integration~\cite{BBCF,BFM,MastroliaTriple} 
have been extended to work within
the dimensional regularization scheme and with massive particles.

In this paper we present the package S@M (Spinors@Mathematica) which implements the spinor-helicity formalism in Mathematica. The package allows the use of complex-spinor algebra along with the multi-purpose features of Mathematica.
The package provides 
\begin{itemize}
\item the definitions of the spinor objects with their basic properties, 
\item functions to manipulate them   
\item numerical evaluation.
\end{itemize} 
These capabilities make the package {\tt S@M} suitable for, for example,
\begin{itemize}
\item[-] the generation of complex spinors associated with solutions 
of multi-particle factorization and of generalized-cut conditions; \\
\item[-] the implementation of BCFW-like recurrence relations
for constructing high-multiplicity tree amplitudes~\cite{BCFW,BGKS}  
and rational coefficients~\cite{SplitHelicityLoop,OnShellOneLoop,Qpap,Bootstrap}; \\
\item[-] the decomposition of massive momenta onto
massless ones, useful for the implementation of the
MHV-rules~\cite{CSW,Risager} and for the BCFW-shift of massive legs \cite{SW}; \\
\item[-] the algebraic manipulation of products of tree amplitudes with
complex spinors sewn in unitarity-cuts. \\
\end{itemize}

The tool presented here is therefore oriented towards
the evaluation of helicity amplitudes at LO and beyond,
which are relevant for the phenomenology of the Standard Model,
for the study of the so called constructible theories~\cite{Benincasa:2007xk}, 
and for the investigation of the ultraviolet-behavior of gravity (see ~\cite{GravityTree,GravityLoop} and references therein).

The paper is organized as follows. Section 2 defines the notation used by the package, Section 3 describes the implementation. In last section, we show three more involved examples of the use of the package {\tt S@M}.     

%----------------------------------------------------------------------------
\section{Notation}
%----------------------------------------------------------------------------
This section describes the conventions and notation used in the package.
%-------------------------------------------------------------------------------
\subsection{Two-dimensional Spinors}
%-------------------------------------------------------------------------------
The two dimensional spinors $\lambda$ and $\tilde \lambda$ for a massless fermion with four momentum $p$ are defined through the Dirac equation
\begin{eqnarray}
\slashed{P}\lambda(p)=0\;, \qquad\tilde \lambda(p) \slashed{P}=0\;,
\end{eqnarray}
with
\begin{eqnarray}
\slashed{P}=p_\mu \sigma^\mu\equiv\left(\begin{array}{cc}p_-&-p_-^\perp\\-p_+^\perp&p_+\end{array}\right),\qquad \sigma^\mu=\left(1,\vec \sigma\right)\;,
\\
\sigma^1=\left(\begin{array}{cc}0&1\\1&0\end{array}\right),\quad
\sigma^2=\left(\begin{array}{cc}0&-i\\i&0\end{array}\right),\quad
\sigma^3=\left(\begin{array}{cc}1&0\\0&-1\end{array}\right)\;.
\end{eqnarray}
The solutions of opposite chirality  are
\begin{eqnarray}
\lambda_a(p)=c\left(\begin{array}{c}p_+\\p_+^\perp\end{array}\right)
&\qquad\mbox{and}\qquad&
\tilde\lambda_{\dot{a}}(p)=\tilde c\left(p_+\quad p_-^\perp\right).
\end{eqnarray}
We define $\underline{\lambda}$ and $\underline{\tilde\lambda}$ as
the spinor with up-indices, obtained by contracting
the spinors $\lambda$ and $\tilde \lambda$ with the 
$\epsilon$-tensor,
\begin{eqnarray}
\underline{\lambda}(p)\equiv\lambda^a(p)
=\epsilon^{ab}\lambda_b(p)
=c\left(p_+^\perp \quad -p_+\right)\,,
&&
\underline{\tilde\lambda}\equiv
\tilde\lambda^{\dot{a}}
=\epsilon^{\dot{a}\dot{b}}\tilde \lambda_{\dot{b}}
=\tilde c\left(\begin{array}{c}p_-^\perp\\-p_+\end{array}\right).
\end{eqnarray}
where $\epsilon^{ab}=\epsilon^{\dot a \dot b}=i\sigma_2$. \\
The underlined notation for spinors carrying up-indices has been introduced here to avoid the introduction of indices in the package {\tt S@M}. 
By making the normalization choice
\begin{equation}
c=\tilde c=\frac{1}{\sqrt{p_+}} \ ,
\end{equation}
we have
\begin{eqnarray} \label{eq:Pdecoinlambda}
\slashed{P}_{\dot{a}a}={\tilde\lambda}_{\dot{a}}{\lambda}_a
\ , 
\qquad
\slashed{P}^{\dot{a}a}=\tilde{\lambda}^{\dot{a}}\lambda^a ,
\end{eqnarray}
so that the 
expressions for the two-dimensional spinors read,
\begin{eqnarray}
\lambda_a(p)=\left(\begin{array}{c}\sqrt{p_+}\\\frac{p_+^\perp}{\sqrt{p_+}}\end{array}\right)\qquad\mbox{and}\qquad
\tilde \lambda_{\dot{a}}(p)=\left(\sqrt{p_+}\,,\quad \frac{p_-^\perp}{\sqrt{p_+}}\right) \ .
\end{eqnarray}
The the above formulas hold as well for 
spinors associated to any
massless complex four-vectors.
For massless real momenta we can 
use the identity,
\begin{equation}\label{eq:lambda}
\sqrt{p_+}\sqrt{p_-}=\sqrt{p_+ p_-}=\sqrt{p_-^\perp p_+^\perp}=\sqrt{p_1^2+p_2^2}\;,
\end{equation}
to write the corresponding spinor,
\begin{eqnarray}\label{eq:twodimsp}
\lambda_a(p)&=&\frac{1}{\sqrt{p_+}}\left(\begin{array}{c}p_+\\p_+^\perp\end{array}\right)=\left(\begin{array}{c}\sqrt{p_+}\\\sqrt{p_-} e^{i\phi}\end{array}\right)\;,
\qquad e^{i\phi}=\frac{p_1+ip_2}{\sqrt{p_1^2+p_2^2}}\;. \\
\tilde{\lambda}_{\dot a}(p) &=&\left(\begin{array}{cc}\sqrt{p_+}\,,&\sqrt{p_-} e^{-i\phi}\end{array}\right) \;.
\end{eqnarray}

%-------------------------------------------------------------------------------
\subsection{Four-dimensional Spinors}
%-------------------------------------------------------------------------------
Positive and negative energy solutions of the four dimensional massless Dirac
equation are identical up to normalization conventions.  
By closely following the definitions of \cite{LanceTASI},
the solutions of definite helicity 
\begin{eqnarray}
u_\pm(k) = \hf(1\pm\gamma_5)u(k)\quad \mbox{ and}\quad v_\mp(k) = \hf(1\pm\gamma_5)v(k) 
\end{eqnarray}
and their conjugates
\begin{eqnarray}
\overline{u_\pm(k)} = \overline{u(k)}\hf(1\mp\gamma_5)\quad \mbox{and} \quad \overline{v_\mp(k)} = \overline{v(k)}\hf(1\mp\gamma_5)\; .
\end{eqnarray}
can be chosen to be equal to each other\footnote{Note that for negative energy solutions, the helicity is the negative of the chirality or $\gamma_5$ eigenvalue}.
For the numerical evaluation of spinor products 
with slashed matrices insertion, we use the Dirac $\gamma$ matrices defined as, 
\be \label{eq:gammamatrices}
\gamma^0\ =\ \left(\begin{array}{cc}{\bf1}&{\bf0}\\{\bf0}&-{\bf1}\end{array}\right)\ ,
\qquad 
\gamma^i\ =\ \left(\begin{array}{cc}{\bf0}&{\bf\sigma}^i\\
                           -{\bf\sigma}^i&{\bf0}\end{array}\right)\ ,
\qquad 
\gamma_5\ =\ \left(\begin{array}{cc}{\bf0}&{\bf1}\\{\bf1}&{\bf0}\end{array}\right)\ ,
\ee
where the entries ${\bf 0}$ and ${\bf 1}$ are $2 \times 2$-matrices. 
In this representation, the massless spinors can be chosen as follows,
\begin{eqnarray}
u_+(k) = v_-(k) = {1\over\sqrt{2}}
  \left[ \begin{array}{c} \lambda_a(p) \\
                  \lambda_a(p)\end{array} \right] , \hskip3mm
u_-(k) = v_+(k) = {1\over\sqrt{2}}
  \left[ \begin{array}{c} 
  \tilde\lambda_{\dot{a}}(p) \\
  -\tilde\lambda_{\dot{a}}(p) 
  \end{array} \right] .
\end{eqnarray}
%\be \label{eq:explicitspinor}
%u_+(k) = v_-(k) = {1\over\sqrt{2}}
%  \left[ \begin{array}{c} \sqrt{k^+} \cr 
%   \sqrt{k^-} e^{i\varphi_k} \\
%                  \sqrt{k^+} \\ 
%   \sqrt{k^-} e^{i\varphi_k} \end{array} \right] , \hskip3mm
%u_-(k) = v_+(k) = {1\over\sqrt{2}}
%  \left[ \begin{array}{c} \sqrt{k^-} e^{-i\varphi_k} \\ 
%                                 -\sqrt{k^+} \\
%                - \sqrt{k^-} e^{-i\varphi_k} \\ 
%                                  \sqrt{k^+}\end{array} \right] , 
%\ee
%where the phase $e^{\pm i\varphi_k}$
%\be \label{eq:phasekdef}
%e^{\pm i\varphi_k}\ \equiv\ 
%  { k^1 \pm ik^2 \over \sqrt{(k^1)^2+(k^2)^2} }
%\ =\  { k^1 \pm ik^2 \over \sqrt{k^+k^-} }\ ,
%\qquad k^\pm\ =\ k^0 \pm k^3.  
%\ee
%\par\noindent
%Plugging Eqs.~\ref{eq:explicitspinor} into the definitions of the spinor
%products, \eqn{basicspinordef}, we get explicit formulae for the case
%when both energies are positive,% {\it i.e.} $k_i^0 >0, \quad k_j^0 > 0$,
%\bea 
%\spa{i}.{j} &=& \sqrt{k_i^- k_j^+} e^{i\varphi_{k_i}}
%              - \sqrt{k_i^+ k_j^-} e^{i\varphi_{k_j}}
%\ =\ \sqrt{|s_{ij}|} e^{i\phi_{ij}}, 
%\nonumber \\ 
%\spb{i}.{j} &=& -\sqrt{k_i^- k_j^+} e^{-i\varphi_{k_i}}
%              + \sqrt{k_i^+ k_j^-} e^{-i\varphi_{k_j}}\;.
%\ =\ \sqrt{|s_{ij}|} e^{-i(\phi_{ij}+\pi)}, 
%\nonumber \\
%\label{eq:posenergyexplicit} 
%&&\qquad\qquad k_i^0 >0, \quad k_j^0 > 0, 
%\eea

\subsubsection{Label Representation}
The spinor-helicity formalism is based on the algebraic manipulation of basic
objects such as the spinor
inner products. They are independent of the explicit representation
used for expressing the spinors themselves. It is useful to have a notation for the spinors that is not bound to a particular explicit representation. In {\tt S@M} we will call this notation the "Label" representation. We define
\begin{eqnarray}\label{eq:spinorshort1}
u_{+}(k_i)\ =\ v_{-}(k_i)\equiv |k_i^{+}\rangle \equiv &\left|i\right> &\equiv \lambda(k_i)\,,  \nonumber\\
u_{-}(k_i)\ =\ v_{+}(k_i)\equiv  |k_i^{-}\rangle\ \equiv&\left|i\right]&\equiv \tilde{\lambda}(k_i)\;,
\end{eqnarray}
and for the conjugate spinors
\begin{eqnarray}\label{eq:spinorshort2}
\overline{u_{+}(k_i)} = \overline{v_{-}(k_i)}\equiv \left[k_i^{+}\right| \equiv &[ i|& \equiv \underline{\lambda}(k_i), \nonumber\\
\overline{u_{-}(k_i)} =\ \overline{v_{+}(k_i)} \equiv  \langle k_i^{-}| \equiv&\left< i\right|&\equiv \underline{\tilde\lambda}(k_i)\;,
\end{eqnarray}
\subsection{Spinor Products}
The scalar (or inner) products for massless spinors are defined as,
\begin{eqnarray}\label{eq:basicspinordef}
\spa{i}.{j}& \equiv& \langle k_i^-|k_j^+\rangle
  \ =\ \overline{u_-(k_i)} u_+(k_j)=\lambda^a(k_i)\lambda_a(k_j)=\underline{\lambda}(k_i)\lambda(k_j),  \\  
\spb{i}.{j} &\equiv& \langle k_i^+|k_j^-\rangle
  \ =\ \overline{u_+(k_i)} u_-(k_j)=\tilde\lambda_{\dot{a}}(k_i)\tilde\lambda^{\dot{a}}(k_j)= \tilde\lambda(k_i)\underline{\tilde\lambda}(k_j). 
\end{eqnarray}
The helicity projection implies that products like 
$[ i | j \rangle$ vanish.

In the rest of the paper we will call \emph{spinor products} 
the $\spa {}.{}$- and $\spb{}.{}$-type spinor inner products, and not the
external spinor-product for the spinorial decomposition of slashed-matrices, 
see Eq.(\ref{eq:Pdecoinlambda}).

The spinor products are, up to a phase, square roots of Lorentz
products, in fact they are related to momenta inner product through 
the identity,
\be \label{eq:spbphasedef}
\spa{i}.{j}\spb{j}.{i}
 = \langle i^- | j^+ \rangle \langle j^+ | i^- \rangle
 = \tr\bigl( \hf(1-\gamma_5) \ksl_i \ksl_j \bigr)
 = 2 k_i\cdot k_j = s_{ij},
\ee
where  $s_{ij}\ =\ (k_i+k_j)^2\ =\ 2 k_i\cdot k_j$.

%The spinor products have simple properties under
%crossing symmetry, as energies become negative \cite{Gunion:1985vca}.  
%For negative energies, we define the spinor product $\spa{i}.{j}$ by analytic continuation
%from the positive energy case, using the same
%formula (\ref{eq:twodimsp}), but with $k_i$ replaced by $-k_i$ if
%$k_i^0 < 0$, and similarly for $k_j$; and with an extra multiplicative 
%factor of $i$ for each negative energy particle. 

%At the level of the solutions $u$ and $v$ of the Dirac equation we have for $k_0<0$
%\begin{equation}
%u_\pm(k)=i u_\pm(-k)\;.
%\end{equation}
We also have the useful identities:
\par\noindent
Gordon identity:
\be \label{eq:spinornorm}
 \langle i | \gamma^\mu | i]\ =
 [i | \gamma^\mu | i\rangle\ =\ 2 k_i^\mu.\qquad\qquad
\ee
Projection operator:  
\begin{equation}\label{eq:spinorproj}
| i \rangle[  i | = \frac{(1+\gamma_5)}{2}\ksl_i \;,      \qquad 
| i ] \langle i | = \frac{1-\gamma_5}{2}\ksl_i      \;,   \qquad 
| i \rangle [ i |+ |i]\langle i | =\ksl_i\;.
\end{equation}
Antisymmetry:
\be \label{eq:spinorantisym}
\spa{j}.{i} = -\spa{i}.{j}, \qquad
\spb{j}.{i} = -\spb{i}.{j}, \qquad
\spa{i}.{i} = \spb{i}.{i} = 0\;. \qquad
\ee
Schouten identity:
\be \label{eq:schouten}
\spa{i}.{j} \spa{k}.{l}\ =\ 
  \spa{i}.{k} \spa{j}.{l} + \spa{i}.{l} \spa{k}.{j} .
\ee
Spinor re-definition:
\bea
 \ksl|i\rangle = |k(i)], && \qquad  
 \langle i| \hspace*{-0.15cm} \ksl = - [ k(i)|,
\nonumber \\
\ksl|i] = |k(i)\rangle,
 && \qquad  
 [i|  \hspace*{-0.15cm} \ksl = - \langle k(i)|.
\eea
\subsection{Massive Spinors Products}

In the spinor-helicity formalism, spinors associated to particles
of momentum $p_I$which are solutions of
the massive Dirac equation, 
$${\overline {u^\pm(p_I)}} \ (\slashed{p}_I - m_I) \ u^\pm(p_I) = 0 \ ,$$ 
can
be as well represented with {\it bra-ket} notation:
\bea
u^+(p_I) &\equiv& |I\rangle \ , \qquad
{\overline {{u}^+(p_I)}} \equiv \langle I | \ , \\
u^-(p_I) &\equiv& |I] \ , \qquad
{\overline {{u}^-(p_I)}} \equiv  [I| \ ,
\eea
where the angle-bracket denote the two spin-polarization states with respect to
a fixed reference axis, rather than helicity states.
Spinor products in this case can be written as \cite{SW,Hall:2007mz} 
%Kleiss-Stirling
\bea
{\overline {{u}^-(p_I)}} u^+(p_J) &\equiv& \spa I.J  \ , \qquad
{\overline {{u}^+(p_I)}} u^-(p_J) \equiv \spb I.J \;,\nonumber \\
{\overline {{u}^+(p_I)}} u^+(p_J)  &\equiv& [I J\rangle \ , \qquad
{\overline {{u}^-(p_I)}} u^-(p_J)   \equiv \langle I J]\;.
\label{eq:def:MassiveSprod}
\eea
One can use the light-cone decomposition of a massive spinor in terms
of two massless one~\cite{Kosower:2004yz}.
Accordingly, by introducing an
arbitrary massless reference momentum $q$,
we can construct a massless momentum ($p_i$)
associated to any massive one ($p_I$),
\bea
p_i^\mu &=& p_I^\mu - {p_I^2 \over 2 p_I\cdot q} q^\mu =
            p_I^\mu - {m_I^2 \over 2 p_i\cdot q} q^\mu \ ,
\qquad p_i^2 = 0 = q^2 \ .
\eea
Correspondingly, the spinor variables read,
\bea
|I\rangle &=& |i\rangle + {m_I \over \spb i.q} |q] \ , \\
|I] &=& |i] + {m_I \over \spa i.q} |q\rangle \ . \\
\eea
With the above formulas we can express the spinor product for
massive particles of Eqs (\ref{eq:def:MassiveSprod}) in terms of the ones involving only massless spinors
\cite{Hall:2007mz} 
\bea
\spa I.J &=& \spa i.j \ , \\
\spb I.J &=& \spb i.j \ , \\
\left[ I \right. J\rangle
&=& \left({m_I \over s_{iq}} + {m_J \over s_{jq}}\right) [i| \slashed{q} | j\rangle \ ,\\
\langle I \left. J\right]
&=& \left({m_I \over s_{iq}} + {m_J \over s_{jq}}\right) 
\langle i| \slashed{q} |j] \ ,
\eea
where $s_{iq} = (p_i + q)^2 = 2 p_i \cdot q = [i q] \langle q i \rangle$.
We notice that in the {\it r.h.s} of the above equations only massless
spinors do appear. Therefore the above definitions of
spinor products associated
to massive particles can have their straightforward implementation in {\tt S@M}.
%%%%%%%%%%%%%%%%%%%%%%%%%%%%%%%%%%%%%%%%%%%%

%------------------------------------------------------------------------
\section{Mathematica Implementation}
%------------------------------------------------------------------------
In the following section we present the Mathematica implementation {\tt S@M}. The Mathematica notebook file {\tt S@M\_Definitions.nb} containing all the examples shown in this section is distributed with the package.

%---------------------------------------------------------------------------
\subsection{Input and Output}
%---------------------------------------------------------------------------
In the notebook, both the input and the output has been designed to look like the familiar bra-ket representation. 
\[\left<\bullet,\bullet\right>, \left[\bullet,\bullet\right], \left<\bullet,\bullet,\bullet\right]\ldots\] 
%The functions of the package can be entered in both the console and the notebook interfaces of Mathematica {\bf using their full name}
%{\bf PM: FullForm ?}. 
In the console version, the functions can only by inputed using their names, in the notebook, several other input methods are available. \begin{itemize}
\item Output from a previous evaluation can be copy-pasted as input for the next evaluation. 
\item Some basic functions and objects like the spinor products are available from a palette that is opened when the package is loaded within the notebook. 
\item Spinor products can be entered using the characters sequnce "ESC~$<$~ESC", "ESC~$>$~ESC","[", "]", "|" if the input format is {\tt TraditionalForm}.  
\end{itemize} 
The examples in the following sections are displayed as they appear in the Mathematica notebook. 

Similar output can be obtained in the console version by using the command {\tt TraditionalForm}. The package is loaded using
\vspace{0.2cm}\\
\fbox{\parbox{\exboxlength}{
\exampleSp\\
\exampleSp
}
}\vspace{0.1cm}\\
The package should be loaded at the beginning of the Mathematica session. 
\subsection{General Structure}
Massless spinor variables are the fundamental objects for the
construction of both inner spinor-products (associated to Lorentz invariants),
and external spinor-products (associated to slashed-matrices). \\
In {\tt S@M}, massless spinors are expected to be {\it labeled} by either symbols or integers. 
The spinor products are independent of the explicit  representation of the spinors. Objects that do not refer to an explicit representation, like spinor products in their $\left<\bullet,\bullet\right>$, $\left[\bullet, \bullet\right]$ form,
are said to be in the "label" representation. To allow more flexibility in the use of the package %{\bf PM: the sentence sounds too obscure}
two explicit spinor representations have been implemented, %{\it i.e.} 
the two-dimensional (Weyl) and the four-dimensional (Dirac) ones. The different representations and the names are summarized in the following table.
\begin{center}
\begin{tabular}{|c|c|c|}
\hline
{Label}&{2 dimensional}&{4 dimensional} \\
%\hline
\hline
%\multicolumn{3}{|c|}{spinors}\\\hline \hline 
$\left|s\right>$,$\left<s\right|$ & $\lambda(k_s)$, $\underline{\lambda}(k_s)$ & $u_+(k_s)$ ,  $\overline{u_-(k_s)}$  \\ 
\hline
 {\tt s} & {\tt La[s]},{\tt CLa[s]} & {\tt USpa[s]}, {\tt UbarSpa[s]} \\
\hline\hline
 $\left|s\right]$,$\left[s\right|$ & 
$\tilde\lambda(k_s)$, $\underline{\tilde\lambda}(k_s)$ & $u_-(k_s)$, $\overline{u_+(k_s)}$ \\ 
\hline
{\tt s} & {\tt Lat[s]},{\tt CLat[s]} & {\tt USpb[s]}, {\tt UbarSpb[s]} \\
\hline
\end{tabular}
\end{center}
The functions provided by the packages act on (defined) spinor-labels, 
and work in  all the representations.
The flexibility of {\tt S@M} relies in the multiple use of 
a given symbol defined as massless spinor, 
which can represent at the same time either the spinor itself or its associated slashed matrix, 
with the automatic understanding of its interpretation according to the context.

\subsection{New Functions}
The new functions introduced by the package are described in the following section. A list of all functions can be found in Appendix \ref{sec:index}.
%_____________________________________________________________________________
\subsubsection{Spinors Declaration}
%
%---------------------------- DeclareSpinor
%
In {\tt S@M} objects called (and declared as) spinors are considered to be 
the solution of the \emph{massless} Dirac equation. 
That is not a restriction on the usability of the package,
since solutions of the massive Dirac equation can be constructed from massless spinors. 
The symbol representing a given spinor is used to represent at the same time, 
and depending on the context, the spinor itself, its vector or the corresponding slashed matrix.  
\begin{description}
\item[\tt DeclareSpinor]\myindex{DeclareSpinor}\rule{0cm}{1cm}\\
The function {\tt DeclareSpinor} can be called with one or a sequence of arguments. It declares its arguments to be spinors. 
If undeclared variables are used as spinors, 
some automatic properties will not be applied and most functions cannot be used. 
\vspace{0.2cm}\\
\rule{-1cm}{0cm}\fbox{\parbox{\exboxlength}{
\exampleSp\\
\exampleSp
}
}\vspace{0.1cm}\\
 Integer labels for spinors do not have to be declared, for more details, see the section on {\tt Sp} below. 
If a symbol is defined as a spinor, it can also be used to represent both its Lorentz vector (see page \pageref{com:DeclareLVector}) or its corresponding slashed matrix.
 
%
%---------------------------- SpinorQ
%
\item[\tt SpinorQ]\myindex{SpinorQ}\rule{0cm}{1cm}\\
{\tt SpinorQ} tests whether its argument has been declared as a spinor or not, it returns {\tt True} if so and {\tt False} otherwise. It can be used for example in patterns.
\vspace{0.2cm}\\
\rule{-1cm}{0cm}\fbox{\parbox{\exboxlength}{
\exampleSp\\
\exampleSp\\
\exampleSp
}
}\vspace{0.1cm}\\
%
%---------------------------- UnDeclareSpinor
%
\item[\tt UndeclareSpinor]\myindex{UndeclareSpinor}\rule{0cm}{1cm}\\
The function {\tt UndeclareSpinor} removes its argument from the list of spinors.
%
%---------------------------- Sp
%
\item[\tt Sp]\myindex{Sp}\rule{0cm}{1cm}\\
Spinors can be labeled by integers using the function {\tt Sp}. The object {\tt Sp[n]} is considered as a spinor. In the scalar products {\tt Spaa}, {\tt Spab}, {\tt Spba} and {\tt Spbb} described below, integer arguments are automatically wrapped into the {\tt Sp} function.
\vspace{0.2cm}\\
\rule{-1cm}{0cm}\fbox{\parbox{\exboxlength}{
\exampleSp\\
\exampleSp\\
\exampleSp\\
\exampleSp
}
}\vspace{0.1cm}\\
In {\tt StandardForm} and {\tt TraditionalForm}, the function {\tt Sp} is not displayed, only its argument.

The function can be made visible by using {\tt FullForm}.

\end{description}
%______________________________________________________________
\subsubsection{Spinor Representations}
\begin{description}
\item[\tt La, Lat, CLa, CLat]\myindex{La}\myindex{Lat}\myindex{CLa}\myindex{CLat}\rule{0cm}{1cm}\\
The two-dimensional representations of the spinor {\tt  s} are given in the following table.
\begin{center}
\begin{tabular}{|c|c|c|c|}
\hline
$\lambda(k_s)$ & $\tilde\lambda(k_s)$ & $\underline{\lambda}(k_s)$ & $\underline{\tilde\lambda}(k_s)$  \\ 
\hline
{\tt La[s]} & {\tt Lat[s]} & {\tt CLa[s]} & {\tt CLat[s]} \\
\hline
\end{tabular}
\end{center}
The arguments of these functions are spinor labels or integers.
In the latter case, the argument, say {\tt i}, is automatically converted to the 
corresponding spinor label, using {\tt Sp[i]} (see above). \\
{\tt La[s]}, {\tt Lat[s]}, {\tt CLa[s]} and {\tt CLat[s]} are linear in their spinor argument. 
\vspace{0.2cm}\\
\rule{-1cm}{0cm}\fbox{\parbox{\exboxlength}{
\exampleSp\\
\exampleSp\\
\exampleSp\\
\exampleSp\\
\exampleSp
}
}\vspace{0.1cm}\\
The contraction of two different two-dimensional spinors is implemented using the Mathematica {\tt Dot} operator
and automatically displayed in a standard order 
(appropriate for the numerical evaluation, see page \pageref{par:StdOrder}). 
\vspace{0.2cm}\\
\rule{-1cm}{0cm}\fbox{\parbox{\exboxlength}{
\exampleSp\\
\exampleSp\\
\exampleSp
}
}\vspace{0.1cm}\\
\item[\tt USpa, USpb, UbarSpa, UbarSpb]\myindex{USpa}\myindex{USpb}\myindex{UbarSpa}\myindex{UbarSpb}\rule{0cm}{1cm}\\
The four-dimensional representation of the spinor {\tt  s} are given in the following table.
\begin{center}
\begin{tabular}{|c|c|c|c|}
\hline
$u_+(k_s)$ & $u_-(k_s)$ &  $\overline{u_-(k_s)}$  & $\overline{u_+(k_s)}$ \\
\hline
{\tt USpa[s]} & {\tt USpb[s]} &  {\tt UbarSpa[s]} & {\tt UbarSpb[s]} \\
\hline
\end{tabular}
\end{center}
These four functions are linear in their spinor argument.
\vspace{0.2cm}\\
\rule{-1cm}{0cm}\fbox{\parbox{\exboxlength}{
\exampleSp\\
\exampleSp\\
\exampleSp\\
\exampleSp\\
\exampleSp
}
}\vspace{0.1cm}\\
The contraction of two different two-dimensional spinors is implemented using the Mathematica {\tt Dot} operator
and automatically displayed in a standard order 
(appropriate for the numerical evaluation, see page \pageref{par:StdOrder}). 
\vspace{0.2cm}\\
\rule{-1cm}{0cm}\fbox{\parbox{\exboxlength}{
\exampleSp\\
\exampleSp\\
\exampleSp
}
}%\vspace{0.1cm}\\
\end{description}
%______________________________________________________________
\subsubsection{Lorentz Vectors}
In {\tt S@M} we call Lorentz vector any 4-dim vector of the form, $k^\mu = (k^0,k^1,k^2,k^3)$. 

%---------------------------- DeclareLVector
%
\begin{description}
\item[\tt DeclareLVector]\myindex{DeclareLVector}\rule{0cm}{1cm}\\
The function {\tt DeclareLVector} can be called with one or a sequence of arguments. It declares its arguments as Lorentz vectors. \\
Momenta associated to spinors (declared through {\tt DeclareSpinor}) do not need to be declared, and
one can use the symbol of the spinor to represent the corresponding Lorentz vector as well. 
To represent the Lorentz vector of a spinor labeled by an integer {\tt i}, one may use {\tt Sp[i]}.
\vspace{0.2cm}\\
\rule{-1cm}{0cm}\fbox{\parbox{\exboxlength}{
\exampleSp\\
\exampleSp
}
}\vspace{0.1cm}\\
%
%---------------------------- UnDeclareLVector
%
\item[\tt UndeclareLVector]\myindex{UndeclareLVector}\rule{0cm}{1cm}\\
The function {\tt UnDeclareLVector} removes its argument from the list of the Lorentz vectors.
\vspace{0.2cm}\\
\rule{-1cm}{0cm}\fbox{\parbox{\exboxlength}{
\exampleSp\\
\exampleSp
}
}\vspace{0.1cm}\\
%
%---------------------------- LVectorQ
%
\item[\tt LVectorQ]\myindex{LVectorQ}\rule{0cm}{1cm}\\
{\tt LVectorQ} tests whether its argument can be interpreted as a Lorentz vector or not. It returns {\tt True} if so and {\tt False} otherwise. It can be used for example in patterns.
\vspace{0.2cm}\\
\rule{-1cm}{0cm}\fbox{\parbox{\exboxlength}{
\exampleSp\\
\exampleSp\\
\exampleSp\\
\exampleSp\\
\exampleSp\\
\exampleSp
}
}\vspace{0.1cm}\\
\end{description}
%_______________________________________________________________________________________________
\subsubsection{Minkowski Products}
%
%---------------------------- MP
%
\begin{description}
\item[\tt {MP[p,q]}]\myindex{MP}\rule{0cm}{1cm}\\
The function {\tt MP[p,q]} represents the Minkowski product 
\[p\cdot q=p^0 q^0-\vect p\cdot \vect q.\]
The two arguments can be either spinors or Lorentz vectors or linear combinations thereof.

Integer arguments are interpreted as the four vectors associated to the spinor {\tt Sp[i]}. {\tt MP} is symmetric in its arguments, so they are automatically sorted. The vector product is linear.
\vspace{0.2cm}\\
\rule{-1cm}{0cm}\fbox{\parbox{\exboxlength}{
\exampleSp\\
\exampleSp\\
\exampleSp
}
}\vspace{0.1cm}\\
\item[\tt {MP2[p]}]\myindex{MP2}\rule{0cm}{1cm}\\
The function {\tt MP2[p]} is a shortcut for {\tt MP[p,p]}.
\vspace{0.2cm}\\
\rule{-1cm}{0cm}\fbox{\parbox{\exboxlength}{
\exampleSp\\
\exampleSp
}
}\vspace{0.1cm}\\
The function {\tt MP} can be used with explicit four vector representations:
\vspace{0.2cm}\\
\rule{-1cm}{0cm}\fbox{\parbox{\exboxlength}{
\exampleSp\\
\exampleSp
}
}
\end{description}
%_______________________________________________________________________________
%_______________________________________________________________________________________________
\subsubsection{Invariants}
%
%---------------------------- sij
%
\begin{description}
\item[\tt s{[i,j]}]\myindex{s}\rule{0cm}{1cm}\\
The function {\tt s[i,j]} represents the kinematic invariant given 
by the square of the sum of two momenta, 
\[s_{ij}=(p_i+p_j)^2.\]
The two arguments can be either spinors or Lorentz vectors. 

Integer arguments are interpreted as the four vectors associated to the spinor {\tt Sp[i]}. Since the scalar product {\tt s} is symmetric in its arguments, they are automatically sorted. 
\vspace{0.2cm}\\
\rule{-1cm}{0cm}\fbox{\parbox{\exboxlength}{
\exampleSp\\
\exampleSp\\
\exampleSp
}
}\vspace{0.1cm}\\
The function {\tt s} also accepts more than two arguments (which can be spinors or Lorentz vectors) for multi-particle invariants,
\[s_{i...j}=(p_i+...+p_j)^2.\]
%\vspace{0.2cm}\\
\rule{-1cm}{0cm}\fbox{\parbox{\exboxlength}{
\exampleSp\\
\exampleSp\\
\exampleSp
}
}
\end{description}
%_______________________________________________________________________________________________

\subsubsection{Slashed Matrices}
Slashed matrices are in general contractions of Lorentz momenta with gamma-matrices $\slashed{P} = P^\mu \gamma_\mu$. 
There are three representations for the slashed matrices in the package, as summarized in the following table. 
\begin{center}
\begin{tabular}{|c|c|c|}
\hline
{Label}&{2 dimensional}&{4 dimensional} \\
\hline\hline
 $\slashed{P}$ & $P^{\dot aa}$, $P_{a\dot a}$ & $\slashed{P}$ \\
\hline
{\tt Sm[P]} & {\tt Sm2[P]}, {\tt CSm2[P]} & {\tt Sm4[P]} \\
\hline\hline
$\left|b\right]\!\!\left<a\right|+\left|a\right>\!\!\left[b\right|$ &
$\underline{\tilde\lambda}(k_b).\underline{\lambda}(k_a) $, $\lambda(k_a).\tilde\lambda(k_b)$ & $\overline{u_+(a)}.u_-(b)+\overline{u_-(b)}.u_+(a) $ \\
\hline
 {\tt SmBA[b,a]} &{\tt SmBA2[b,a]}, {\tt CSmBA2[b,a]} & {\tt SmBA4[b,a]} \\
\hline
\end{tabular}
\end{center}

%In {\tt S@M } 
%there are four possible way of generating slashed matrices, 
%eventually according to the basic variables
%used to define them:
%\begin{itemize}
%\item simply declared slashed matrix, 
%      without specifying the link to other variables (see {\tt DeclareSMatrix});
%\item associated to a declared massless spinor, say $a$:
%      $ \slashed{a} = |a]\langle a| + |a\rangle [a| \ $ (see {\tt Sm}); 
%\item associated to a declared L-vector, say $P^\mu$:
%      $ \slashed{P} = P^\mu \gamma_\mu \ $ (see {\tt Sm}) ;
%\item generated by the tensor product of two massless spinors, say $a$ and $b$:
%      $ \slashed{\epsilon} =  |a\rangle [b| + |b]\langle a| $ (see {\tt SmBA});
%\end{itemize}
In {\tt S@M} slashed matrices are used either inside spinor products or as explicit matrices. There are four different types of slashed matrices.
\begin{itemize}
\item slashed matrices corresponding to a declared (massless) spinor. These slashed matrices are 
labeled with the same symbol as their spinor. 
The slashed matrix associated with the massless spinor {\tt a} are represented by:
 {\tt Sm[a]}, in the label representation; {\tt Sm2[a]} or {\tt CSm2[a]}, in the two-dimensional representation;
 and {\tt Sm4[a]}, in the four-dimensional representation. 
Inside spinor products in the label representation, the function {\tt Sm} is not necessary and can be omitted. 
\item %tensor 
external products of spinors. 
These slashed matrices are represented by the functions: {\tt SmBA}, in the label representation; 
{\tt SmBA2} or {\tt CSmBA2}, in the two dimensional representation; and {\tt SmBA4}, 
in the four-dimensional representation.     
\item slashed matrices corresponding to a declared (possibly massive) vector. 
These slashed matrices are labeled with the same symbol as their vector. 
The slashed matrix associated with the vector {\tt P} are represented by:
 {\tt Sm[P]} in the label representation; {\tt Sm2[P]} or {\tt CSm2[P]}, in the two dimensional representation;
 and {\tt Sm4[p]}, in the four-dimensional representation. 
Inside spinor products in the label representation, the function {\tt Sm} is not necessary and can be omitted. 
\item other slashed matrices unrelated to a declared vector or spinor. These slashed matrices have to be declared. The slashed matrix declared with the symbol {\tt Q} is represented by {\tt Sm[Q]} in the label representation, {\tt Sm2[Q]} or {\tt CSm2[Q]}, in the two dimensional representation and {\tt Sm4[Q]}, in the four-dimensional representation.
Inside spinor products in the label representation, the function {\tt Sm} is not necessary and can be omitted. 
\end{itemize} 
\begin{description}
%---------------------------- DeclareSMatrix
%
\item[\tt DeclareSMatrix]\rule{0cm}{1cm}\\
The function {\tt DeclareSMatrix} can be called with one or a sequence of arguments. 
It declares its arguments to be slashed matrices. 
Slashed matrices corresponding to declared spinors and Lorentz vectors are 
are labeled by the same symbol as the spinor or Lorentz vector and thus do not need to be re-defined.
If undeclared variables are used as slashed matrices, 
some automatic properties will not be applied, and most functions cannot be used. 
\vspace{0.2cm}\\
\rule{-1cm}{0cm}\fbox{\parbox{\exboxlength}{
\exampleSp\\
\exampleSp
}
}\vspace{0.1cm}\\
%
%---------------------------- UndeclareDMatrix
%
\item[\tt UndeclareSMatrix]\myindex{UndeclareSMatrix}\rule{0cm}{1cm}\\
The function {\tt UndeclareSMatrix} removes its argument from the list of the slashed matrices.
%
%---------------------------- SMatrixQ
%
\item[\tt SMatrixQ]\myindex{SMatrixQ}\rule{0cm}{1cm}\\
{\tt SMatrixQ} tests whether its argument has been declared as a slashed matrix or not, it returns {\tt True} if so and {\tt False} otherwise. It can be used for example in patterns.
\vspace{0.2cm}\\
\rule{-1cm}{0cm}\fbox{\parbox{\exboxlength}{
\exampleSp\\
\exampleSp\\
\exampleSp
}
}\vspace{0.1cm}\\
\item[\tt SmBA]\myindex{SmBA}\rule{0cm}{1cm}\\
The object {\tt SmBA[b,a]} represents 
slashed matrices formed by the tensor product of two spinors, like 
\[\left|b\right]\!\!\left<a\right|+\left|a\right>\!\!\left[b\right|. \]
%When inserted in a spinor chain, one of the terms always vanishes. 
The arguments {\tt a} and {\tt b} are spinors labels.
{\tt SmBA} is linear in both arguments. 
If the two arguments are equal, {\tt SmBA[a,a]} is automatically replaced by {\tt Sm[a]}.
\vspace{0.2cm}\\
\rule{-1cm}{0cm}\fbox{\parbox{\exboxlength}{
\exampleSp\\
\exampleSp\\
\exampleSp\\
\exampleSp
}
}
%\vspace{0.1cm}\\

\end{description}
%______________________________________________________________________
\subsubsection{Slashed Matrices Representations}
\begin{description}

%---------------------------- Sm
%
\item[\tt Sm]\myindex{Sm}\rule{0cm}{1cm}\\
The object {\tt Sm} is used for slashed matrices corresponding 
to previously declared spinors and Lorentz vectors. 
{\tt Sm} can be called with one argument, being either a spinor label or a vector label.
In particular, slashed matrices associated either 
to spinors (declared through {\tt DeclareSpinor}),
or to vectors (declared through {\tt DeclareLVector}) are automatically declared. 
One can use the symbol of the spinor, say {\tt s}, or the one of the vector, say {\tt P} 
to represent the corresponding slashed matrix by means of {\tt Sm[s]} or {\tt Sm[P]} respectively.

The object {\tt Sm} is linear in its argument.
\vspace{0.2cm}\\
\rule{-1cm}{0cm}\fbox{\parbox{\exboxlength}{
\exampleSp\\
\exampleSp\\
\exampleSp
}
}\vspace{0.1cm}\\
\item[\tt Sm2, CSm2]\myindex{Sm2}\myindex{CSm2}\rule{0cm}{1cm}\\
The two-dimensional representations of the slashed matrix {\tt P} are given in the following table.
\begin{center}
\begin{tabular}{|c|c|}
\hline
$P^{\dot aa}$ & $P_{a\dot a}$  \\ 
\hline
{\tt Sm2[P]} & {\tt CSm2[P]}  \\
\hline
\end{tabular}
\end{center}
The functions {\tt Sm2} and {\tt CSm2} are linear in their spinor argument. The contraction with another slashed matrix or with a spinor is implemented using the Mathematica {\tt Dot} operator. 
\vspace{0.2cm}\\
\rule{-1cm}{0cm}\fbox{\parbox{\exboxlength}{
\exampleSp\\
\exampleSp\\
\exampleSp
}
}
\item[\tt Sm4]\myindex{Sm4}\rule{0cm}{1cm}\\
The four-dimensional representations of the slashed matrix {\tt P } is given by {\tt Sm4[P]}. This function is linear in its spinor argument. The contraction of two different four-dimensional matrices or with four-dimensional spinors is implemented using the Mathematica {\tt Dot} operator.
\vspace{0.2cm}\\
\rule{-1cm}{0cm}\fbox{\parbox{\exboxlength}{
\exampleSp\\
\exampleSp\\
\exampleSp
}
}
\item[\tt SmBA2]\myindex{SmBA2}\myindex{CSmBA2}\rule{0cm}{1cm}\\
In the two-dimensional representation, the slashed matrices build from two spinors 
\[\underline{\tilde\lambda}^{\dot b}.\underline{\lambda}^a \qquad \mbox{and}\qquad \lambda_a.\tilde\lambda_{\dot b} \]
are represented by the two function {\tt SmBA2[b,a]}, {\tt CSmBA2[b,a]} respectively. These functions are linear in both spinor arguments. 
\vspace{0.2cm}\\
\rule{-1cm}{0cm}\fbox{\parbox{\exboxlength}{
\exampleSp\\
\exampleSp\\
\exampleSp
}
}\vspace{0.1cm}\\
Their multiplication with other slashed matrices are implemented using the Mathematica {\tt Dot} operator. {\tt SmBA2} and {\tt CSmBA2} are replaced by their tensor product representation when they are inserted in a spinor chain.
\vspace{0.2cm}\\
\rule{-1cm}{0cm}\fbox{\parbox{\exboxlength}{
\exampleSp\\
\exampleSp\\
\exampleSp
}
}
\item[\tt SmBA4]\myindex{SmBA4}\rule{0cm}{1cm}\\
The slashed matrix build from two spinors, 
\[\left|b\right]\!\!\left<a\right|+\left|b\right]\!\!\left<a\right|\;,\]
is represented in the four-dimensional representation by the function {\tt SmBA4[b,a]}. This function is linear in both spinor arguments.
\vspace{0.2cm}\\
\rule{-1cm}{0cm}\fbox{\parbox{\exboxlength}{
\exampleSp\\
\exampleSp
}
}\vspace{0.1cm}\\
Its multiplication with other slashed matrices is implemented using the Mathematica {\tt Dot} operator.  {\tt SmBA4} is replaced by its tensor product representation when inserted in a spinor chain.
\vspace{0.2cm}\\
\rule{-1cm}{0cm}\fbox{\parbox{\exboxlength}{
\exampleSp\\
\exampleSp\\
\exampleSp
}
}
\end{description}
%___________________________________________________

%_______________________________________________________________________________
\subsubsection{Spinor Products}

\myindex{Spaa}\myindex{Spbb}\myindex{Spab}\myindex{Spba}
Spinor products are represented in {\tt S@M} by four different objects: {\tt Spaa}, {\tt Spab}, {\tt Spba} and {\tt Spbb}, according to the following table. 
\begin{center}
\begin{tabular}{|c|c|c|c|}
\hline
$\left<a ...b\right>$ & $\left<a...b\right]$ & $\left[a... b\right>$ & $\left[a... b\right]$\\\hline
{\tt Spaa[a,...,b]} &{\tt Spab[a,...,b]} & {\tt Spba[a,...,b]} & {\tt Spbb[a,...,b]} \\\hline
\end{tabular}
\end{center}
The left- and right-most arguments are spinors and the intermediate arguments are slashed matrices or objects that can be interpreted as slashed matrices such as spinors or Lorentz vectors. {\tt Spaa} and {\tt Spbb} expect an even number of (declared) arguments whereas {\tt Spab} and {\tt Spba} expect an odd number of (declared) arguments. If not all the arguments are integers (and therefore automatically interpreted as spinors) or declared either as spinors, Lorentz vectors or slashed matrices, the properties listed below cannot be applied.    
\begin{description}
\item[-Standard order]\rule{2cm}{0cm}\label{par:StdOrder}\\
The spinor products have a normal ordering for their arguments. If the rightmost and leftmost elements are spinors, the middle elements are slashed matrices (or can be interpreted as such) and in addition if the spinors are not in the standard order, %they are brought into it 
the spinor product is ordered using the identities 
\begin{eqnarray*}
\left<b\,a\right>=-\left<a\,b\right>,&\qquad&\left[b\,a\right]=-\left[a\,b\right],\\
\left<b|Q...P|a\right>=-\left<a|P...Q|b\right>,&\qquad&\left[b|Q...P|a\right]=-\left[a|P...Q|b\right],\\
\left[b|P|a\right>=\left<a|P|b\right],&\qquad&\left[b|Q...P|a\right>=\left<	a|P...Q|b\right].
\end{eqnarray*}

A special case of these identities is the on-shell condition 
\[\left<a\,a\right>=0,\quad  \left[a\,a\right]=0\;.\]
%and therefore
%\[\left<...a\,a ...\right>=0,\quad  \left[...a\,a...\right]=0,\quad  \left<...a\,a...\right]=0,\quad  \left[...a\,a...\right>=0.\]
\rule{-1.85cm}{0cm}\fbox{\parbox{\exboxlength}{
\exampleSp\\
\exampleSp\\
\exampleSp\\
\exampleSp\\
\exampleSp\\
\exampleSp
}
}\vspace{0.1cm}\rule{2cm}{0cm}\\
The normal ordering of the {\tt Spaa} and {\tt Spbb} products are opposite so that the products $\left<a\,b\right>\left[b\,a\right]$ are displayed in this usual way.

The same rules apply to the spinor products written in the two- and four dimensional representations.
\vspace{0.2cm}\\
\rule{-1.85cm}{0cm}\fbox{\parbox{\exboxlength}{
\exampleSp\\
\exampleSp\\
\exampleSp\\
\exampleSp\\
\exampleSp\\
\exampleSp\\
\exampleSp\\
\exampleSp\\
\exampleSp\\
\exampleSp\\
\exampleSp\\
\exampleSp
}
}\vspace{0.1cm}\rule{2cm}{0cm}\\
 
\item[-Linearity] \rule{2cm}{0cm}\\
If the arguments of  a spinor product have been defined as spinors using {\tt DeclareSpinor} or {\tt DeclareSMatrix}, then linear combinations of spinors are automatically expanded. 
\vspace{0.2cm}\\
\rule{-1.85cm}{0cm}\fbox{\parbox{\exboxlength}{
\exampleSp\\
\exampleSp\\
\exampleSp\\
\exampleSp\\
\exampleSp\\
\exampleSp\\
\exampleSp
}
}\vspace{0.1cm}\\
The linearity works for integer labels too, but one has to write the {\tt Sp} function explicitly, since the sum or product of the integers is done before 
wrapping the result with the function {\tt Sp}. 
\vspace{0.2cm}\\
\rule{-1.85cm}{0cm}\fbox{\parbox{\exboxlength}{
\exampleSp\\
\exampleSp\\
\exampleSp\\
\exampleSp
}
}\vspace{0.1cm}\\
\item[-Syntax correction] \rule{2cm}{0cm}\\
When all the arguments of a spinor product are declared either as spinor, Lorentz vector or slashed matrix and their number does not match the expected number for the particular spinor product type (even for {\tt Spaa}, {\tt Spbb}, for odd for {\tt Spab}, {\tt Spba}), the type of the spinor product is changed automatically and a warning is issued. 
\vspace{0.2cm}\\
\rule{-1.85cm}{0cm}\fbox{\parbox{\exboxlength}{
\exampleSp\\
\exampleSp\\
\exampleSp\\
\exampleSp\\
\exampleSp\\
\exampleSp\\
\exampleSp\\
\exampleSp\\
\exampleSp
}
}\vspace{0.1cm}\\
\item[-Slashed matrices insertion] \rule{2cm}{0cm}\\
The Dirac equation and on-shell condition
\[ \slashed{a}\left|a\right>=0,\quad \slashed{a}\left|a\right]=0,\quad \slashed{a}\slashed{a}=a^2=0, \]
are used for spinors used as slashed matrices. The inserted {\tt SmBA} objects are automatically expanded, as shown in the following examples. \vspace{0.2cm}\\
\rule{-1.85cm}{0cm}\fbox{\parbox{\exboxlength}{
\exampleSp\\
\exampleSp\\
\exampleSp\\
\exampleSp\\
\exampleSp\\
\exampleSp\\
\exampleSp\\
\exampleSp\\
\exampleSp\\
\exampleSp\\
\exampleSp\\
\exampleSp\\
\exampleSp\\
\exampleSp\\
\exampleSp
}
}\vspace{0.1cm}\\
\end{description}
%_______________________________________________________________________________________________
\subsubsection{Spinor Manipulations}
\begin{description}
%
%----------------------------  ExpandSToSpinors ConvertSpinorsToS
%
\item[\tt ExpandSToSpinors, ConvertSpinorsToS]\myindex{ExpandSToSpinors}\myindex{ConvertSpinorsToS}\rule{2cm}{0cm}\\
The function {\tt ExpandSToSpinors}, {\tt ConvertSpinorsToS} convert invariants {\tt s} to spinor products and conversely. 
\vspace{0.2cm}\\
\rule{-1cm}{0cm}\fbox{\parbox{\exboxlength}{
\exampleSp\\
\exampleSp\\
\exampleSp\\
\exampleSp\\
\exampleSp\\
\exampleSp\\
\exampleSp\\
\exampleSp\\
\exampleSp\\
\exampleSp
}
}\vspace{0.1cm}\\
%
%----------------------------  SpOpen
%
\item[\tt SpOpen, SpClose]\myindex{SpOpen, SpClose}\rule{2cm}{0cm}\\
The function {\tt SpOpen} decomposes spinor chains containing any slashed matrix that corresponds to a massless spinor 
with products of smaller spinor chains, by applying the definition of such a matrix in terms of its
opposite-chirality spinors,
$$
\slashed{k} = |k]\langle k| + |k\rangle[k| \ .
$$
The function {\tt SpClose} has the reverse effect as that of {\tt SpOpen}. 
It attempts to replace products of spinor products with spinor chains containing slashed matrices.\\
Both the functions can take either one or two arguments. 
The first argument is the expression to be manipulated; the second argument must be a spinor.
With one argument, the functions open or close wherever possible.  
\vspace{0.2cm}\\
\rule{-1cm}{0cm}\fbox{\parbox{\exboxlength}{
\exampleSp\\
\exampleSp\\
\exampleSp\\
\exampleSp\\
\exampleSp\\
\exampleSp\\
\exampleSp\\
\exampleSp\\
\exampleSp\\
\exampleSp\\
\exampleSp\\
\exampleSp\\
\exampleSp\\
\exampleSp\\
\exampleSp\\
\exampleSp\\
\exampleSp
}
}\vspace{0.1cm}\\
If there are different possibilities of reconstructing the spinor chain, {\tt SpClose} does not search for the longest possible spinor chain. The result will depend on the ordering of the spinor labels and might not be invariant under relabeling of the spinor labels.
\vspace{0.2cm}\\
\rule{-1cm}{0cm}\fbox{\parbox{\exboxlength}{
\exampleSp\\
\exampleSp\\
\exampleSp
}
}\vspace{0.1cm}\\
If a spinor is given as a second argument, 
{\tt SpOpen} and {\tt SpClose} will only open or close spinor 
chains containing this specified spinor.
\vspace{0.2cm}\\
\rule{-1cm}{0cm}\fbox{\parbox{\exboxlength}{
\exampleSp\\
\exampleSp\\
\exampleSp\\
\exampleSp\\
\exampleSp\\
\exampleSp\\
\exampleSp\\
\exampleSp\\
\exampleSp\\
\exampleSp\\
\exampleSp\\
\exampleSp
}
}\vspace{0.1cm}\\

%
%----------------------------  To2DimSpinor
%
\item[\tt To2DimSpinor]\myindex{To2DimSpinor}\rule{2cm}{0cm}\\
The function {\tt To2DimSpinor} converts spinor products in the label representation into the two-dimensional representation.
\vspace{0.2cm}\\
\rule{-1cm}{0cm}\fbox{\parbox{\exboxlength}{
\exampleSp\\
\exampleSp\\
\exampleSp\\
\exampleSp
}
}\vspace{0.1cm}\\
There is an ambiguity in the conversion of explicit slashed matrices (see {\tt Sm}), when they are not embedded in a spinor chain. In those cases {\tt To2DimSpinor} convert the slashed matrices to the {\tt Sm2} type. In case of {\tt Dot} products of slashed matrices, the product is transformed to a {\tt Dot} product starting with a {\tt Sm2} matrix.
\vspace{0.2cm}\\
\rule{-1cm}{0cm}\fbox{\parbox{\exboxlength}{
\exampleSp\\
\exampleSp\\
\exampleSp\\
\exampleSp
}
}\vspace{0.1cm}\\
%
%----------------------------  To4DimSpinor
%
\item[\tt To4DimSpinor]\myindex{To4DimSpinor}\rule{2cm}{0cm}\\
The function {\tt To4DimSpinor} converts spinor products in the label representation into the four-dimensional representation.
\vspace{0.2cm}\\
\rule{-1cm}{0cm}\fbox{\parbox{\exboxlength}{
\exampleSp\\
\exampleSp\\
\exampleSp\\
\exampleSp\\
\exampleSp\\
\exampleSp
}
}\vspace{0.1cm}\\
%
%----------------------------  ToSpinorLabel
%
\item[\tt ToSpinorLabel]\myindex{ToSpinorLabel}\rule{2cm}{0cm}\\
The function {\tt To2SpinorLabel} converts spinor products in the two- or four-dimensional representation into the label notation.
\vspace{0.2cm}\\
\rule{-1cm}{0cm}\fbox{\parbox{\exboxlength}{
\exampleSp\\
\exampleSp\\
\exampleSp\\
\exampleSp\\
\exampleSp\\
\exampleSp\\
\exampleSp\\
\exampleSp\\
\exampleSp\\
\exampleSp\\
\exampleSp\\
\exampleSp\\
\exampleSp\\
\exampleSp
}
}\vspace{0.1cm}\\
%----------------------------  Compactify
%
\item[\tt Compactify]\myindex{Compactify}\myindex{ACompactify}\myindex{BCompactify}\rule{2cm}{0cm}\\
Given a slashed matrix, say 
$\slashed{P}$, and a spinor of the $ |\bullet\rangle $-type, say $|b \rangle$,
one can construct the massless spinor
\[\slashed{P}\left|b\right>\;,\]
which indeed is of the opposite chirality $|\bullet]$-type. 
In fact, one can define, 
\[\slashed{P}\left|b\right>\equiv\left|P(b)\right],\quad \left<b\right|\slashed{P}\equiv-\left[P(b)\right| \ ,\]
and, similarly,
\[\slashed{P}\left|b\right]\equiv\left|P(b)\right>,\quad \left[b\right|\slashed{P}\equiv-\left<P(b)\right|\;.\]
In the package, the notation used for the spinor obtained by applying the 
slashed matrix to the spinor {\tt a} is 
{\tt P[a]}.
\vspace{0.2cm}\\
\rule{-1cm}{0cm}\fbox{\parbox{\exboxlength}{
\exampleSp\\
\exampleSp\\
\exampleSp
}
}\vspace{0.1cm}\\
 These objects are recognized as spinors. The application of more than one slashed matrix to a spinor is done using the Mathematica {\tt Dot} operator. The function {\tt Compactify} uses this definition 
to reduce spinor products with inserted slashed matrices to spinor products of two spinor objects.
\vspace{0.2cm}\\
\rule{-1cm}{0cm}\fbox{\parbox{\exboxlength}{
\exampleSp\\
\exampleSp\\
\exampleSp\\
\exampleSp\\
\exampleSp
}
}\vspace{0.1cm}\\
One can specify a spinor as a second argument for {\tt Compactify}. In this case the spinor products containing the given spinor are compactified in such a way that the specified spinor is left untouched. 
\vspace{0.2cm}\\
\rule{-1cm}{0cm}\fbox{\parbox{\exboxlength}{
\exampleSp\\
\exampleSp\\
\exampleSp\\
\exampleSp\\
\exampleSp
}
}\vspace{0.1cm}\\
The functions {\tt ACompactify[x,a]} and {\tt BCompactify[x,a]} work the same way as {\tt Compactify[x,a]} but only spinor products containing $\left[a\right>$ or $\left|a\right]$ respectively are compactified.  
\vspace{0.2cm}\\
\rule{-1cm}{0cm}\fbox{\parbox{\exboxlength}{
\exampleSp\\
\exampleSp\\
\exampleSp\\
\exampleSp
}
}\vspace{0.1cm}\\
%
%----------------------------  Uncompact
%
\item[\tt UnCompact]\myindex{UnCompact}\rule{2cm}{0cm}\\
The function {\tt UnCompact} uncompactifies the spinor products compactified with {\tt Compactify}.
\vspace{0.2cm}\\
\rule{-1cm}{0cm}\fbox{\parbox{\exboxlength}{
\exampleSp\\
\exampleSp\\
\exampleSp\\
\exampleSp\\
\exampleSp\\
\exampleSp\\
\exampleSp
}
}\vspace{0.1cm}\\
One can specify a spinor as a second argument. In this case only the spinor products where the Dirac matrices are compactified onto the specified spinor will be uncompactified. 
\vspace{0.2cm}\\
\rule{-1cm}{0cm}\fbox{\parbox{\exboxlength}{
\exampleSp\\
\exampleSp\\
\exampleSp\\
\exampleSp\\
\exampleSp
}
}\vspace{0.1cm}\\
%
%----------------------------  Schouten
%
\item[\tt Schouten]\myindex{Schouten}\rule{2cm}{0cm}\\
The function {\tt Schouten} applies the Schouten identities
$$
\spa i.j \spa k.\ell
= \spa i.\ell \spa k.j + \spa i.k \spa j.\ell \ ,
\qquad
\spb i.j \spb k.\ell
= \spb i.\ell \spb k.j + \spb i.k \spb j.\ell \ .
$$
There are three different applications of the function.
\begin{description}
\item[\tt Schouten{[x,i,j,k,l]}]\rule{2cm}{0cm}\\
The function with four spinor arguments will search {\tt x} for occurrences of the products $\spa i.j \spa k.\ell$ or $\spb i.j \spb k.\ell$ and replace it using the above identities. 
\vspace{0.2cm}\\
\rule{-1.85cm}{0cm}\fbox{\parbox{\exboxlength}{
\exampleSp\\
\exampleSp\\
\exampleSp\\
\exampleSp\\
\exampleSp\\
\exampleSp
}
}\vspace{0.1cm}\\
\item[\tt Schouten{[x,i,j,k]}]\rule{2cm}{0cm}\\
The function with three spinor arguments will search for occurrences of the spinor  product $\left<i\, j\right>$ or $\left[i\, j\right]$ and will try to use the Schouten identity to combine it with the spinor {\tt k}. 
\vspace{0.2cm}\\
\rule{-1.85cm}{0cm}\fbox{\parbox{\exboxlength}{
\exampleSp\\
\exampleSp\\
\exampleSp\\
\exampleSp
}
}\vspace{0.1cm}\\
\item[\tt Schouten{[x,l]}]\rule{2cm}{0cm}\\
The function with one spinor arguments will search for 
structures like
$$
{ \spa \ell.u \over \spa \ell.s \spa \ell.t} \ ,
\qquad
{ \spb \ell.u \over \spb \ell.s \spb \ell.t} \ ,
$$
and will use the Schouten identities to split them into partial fractions, 
$$
{ \spa \ell.u \over \spa \ell.s \spa \ell.t} 
=   {\spa s.u \over \spa \ell.s \spa s.t} 
  - {\spa t.u \over \spa \ell.t \spa s.t} \ , 
\qquad 
{ \spb \ell.u \over \spb \ell.s \spb \ell.t} 
=   {\spb s.u \over \spb \ell.s \spb s.t} 
  - {\spb t.u \over \spb \ell.t \spb s.t}\,.
$$
The function also works for spinor products with embedded Dirac matrices.
\vspace{0.2cm}\\
\rule{-1.85cm}{0cm}\fbox{\parbox{\exboxlength}{
\exampleSp\\
\exampleSp\\
\exampleSp
}
}\vspace{0.1cm}\\
\item[\tt ASchouten, BSchouten]\myindex{ASchouten}\myindex{BSchouten}\rule{2cm}{0cm}\\
The function {\tt ASchouten} and {\tt BSchouten} behave like {\tt Schouten} 
but apply the Schouten identity selectively on  
$\left| \ell \right>$-variables and $\left| \ell \right]$-variables 
respectively.     
\vspace{0.2cm}\\
\rule{-1.85cm}{0cm}\fbox{\parbox{\exboxlength}{
\exampleSp\\
\exampleSp\\
\exampleSp\\
\exampleSp\\
\exampleSp\\
\exampleSp
}
}\vspace{0.1cm}\\
\end{description}
\item[\tt ASpinorReplace{[x,a,n]},BSpinorReplace{[x,a,n]}]\myindex{ASpinorReplace}\myindex{BSpinorReplace}\rule{2cm}{0cm}\\
The functions {\tt ASpinorReplace} and {\tt BSpinorReplace} replace the spinor  {\tt a} in expression {\tt x} with {\tt n}. {\tt ASpinorReplace} only replaces the spinors $\left|a\right>$ and {\tt BSpinorReplace} replaces only $\left|a\right]$. Slashed matrices corresponding to the spinor {\tt a} will be split according to 
\[\slashed{a}=\left| a\right>\left[a\right|+\left| a\right]\left<a\right|\]
and the appropriate component will be replaced.  
\vspace{0.2cm}\\
\rule{-1.0cm}{0cm}\fbox{\parbox{\exboxlength}{
\exampleSp\\
\exampleSp\\
\exampleSp\\
\exampleSp\\
\exampleSp\\
\exampleSp\\
\exampleSp\\
\exampleSp
}
}\vspace{0.1cm}\\
\item[\tt ASpinorShift{[x,a,s]},BSpinorShift{[x,a,s]}]\myindex{ASpinorShift}\myindex{BSpinorShift}\rule{2cm}{0cm}\\
The functions {\tt ASpinorShift} and {\tt BSpinorShift} shift the spinor variable {\tt a} in expression {\tt x} with {\tt s}. {\tt ASpinorShift} only shifts the spinor $\left|a \right>$ and {\tt BSpinorShift} only shifts the  spinors $\left|a\right]$. Slashed matrices corresponding to the spinor {\tt a} will be split according to 
\[\slashed{a}=\left| a\right>\left[a\right|+\left| a\right]\left<a\right|\]
and the appropriate component will be shifted. 
\vspace{0.2cm}\\
\rule{-1.0cm}{0cm}\fbox{\parbox{\exboxlength}{
\exampleSp\\
\exampleSp\\
\exampleSp\\
\exampleSp\\
\exampleSp\\
\exampleSp
}
}\vspace{0.1cm}\\
The shift parameter {\tt s} will be interpreted as a spinor of the appropriate chirality. It can be a sum of spinors.
\vspace{0.2cm}\\
\rule{-1.0cm}{0cm}\fbox{\parbox{\exboxlength}{
\exampleSp\\
\exampleSp
}
}\vspace{0.1cm}\\
To account for more generic spinor definitions and shifts composite-spinors may be required. We can use $$
{\tt a, \ \  P[b], \ \ (P.Q)[c], \ \ (P.Q.R)[d],} \ \ \ldots
$$
as the shift argument of the {\tt Shift}-functions to represent the compactified objects like
$$
|a\rangle, \ \ |P(b)\rangle, \ \ |P.Q(c)\rangle, \ \ |P.Q.R(d)\rangle, \ \ \ldots
$$
in the case of an {\tt A}-shift. The expressions obtained using this kind of arguments can be later un-compactified by using {\tt UnCompact}.
\vspace{0.2cm}\\
\rule{-1.0cm}{0cm}\fbox{\parbox{\exboxlength}{
\exampleSp\\
\exampleSp\\
\exampleSp\\
\exampleSp\\
\exampleSp\\
\exampleSp\\
\exampleSp
}
}\vspace{0.1cm}\\
 
\item[\tt ShiftBA{[b,a,z][x]}]\myindex{ShiftBA}\rule{2cm}{0cm}\\
The function {\tt ShiftBA[b,a,z]} performs the shifts combination 
\begin{eqnarray}
\left|b\right]\to \left|b\right]-z  \left|a\right],
\qquad
\left|a\right>\to \left|a\right>+z  \left|b\right>.
\end{eqnarray}
The arguments {\tt a} and {\tt b} must be declared as spinors.
\vspace{0.2cm}\\
\rule{-1.0cm}{0cm}\fbox{\parbox{\exboxlength}{
\exampleSp\\
\exampleSp\\
\exampleSp\\
\exampleSp\\
\exampleSp\\
\exampleSp
}
}\vspace{0.1cm}\\

\end{description}
%%%%%%%%%%%%%%%%%%%%%%%%%%%%%%%%%%%%%%%%%%%%%%%
\subsubsection{Constants}
\myindex{Gamma0}\myindex{Gamma1}\myindex{Gamma2}\myindex{Gamma3}\myindex{Gamma5}
{\tt Gamma0, Gamma1, Gamma2, Gamma3, Gamma5} are the $\gamma$-matrices in the representation (\ref{eq:gammamatrices}).
\vspace{0.2cm}\\
\rule{-1cm}{0cm}\fbox{\parbox{\exboxlength}{
\exampleSp\\
\exampleSp\\
\exampleSp
}
}\vspace{0.1cm}\\
The $\gamma$-matrices can be used as slashed matrices too.
\vspace{0.2cm}\\
\rule{-1cm}{0cm}\fbox{\parbox{\exboxlength}{
\exampleSp\\
\exampleSp
}
}\vspace{0.1cm}\\
\myindex{ProjPlus}\myindex{ProjMinus}
{\tt ProjPlus, ProjMinus} are the helicity projectors $(1\pm\gamma_5)/2$.
\vspace{0.2cm}\\
\rule{-1cm}{0cm}\fbox{\parbox{\exboxlength}{
\exampleSp\\
\exampleSp\\
\exampleSp
}
}\vspace{0.1cm}\\
%
%---------------------------- $SpinorFunctions 
%
\myindex{$SpinorFunctions}
The symbol {\tt \$SpinorFunctions} contains a list of all functions of the package.
\vspace{0.2cm}\\
\rule{-1cm}{0cm}\fbox{\parbox{\exboxlength}{
\exampleSp\\
\exampleSp
}
}%\vspace{0.1cm}\\
\subsubsection{Numerics}
%%%%%%%%%%%%%%%%%%%%%%%%%%  NUMERICS  %%%%%%%%%%%%%%%%%%%%%%%%%%%%%%%%%
The spinor products have a numerical implementation. The first step to use this numerical implementation is to define or generate the four momenta for the spinors. For this there are several possibilities described in the following section.
\subsection{Momentum Generation}
\begin{description}
\item[\tt DeclareSpinorMomentum]\myindex{DeclareSpinorMomentum}\rule{1cm}{0cm}\\
This function takes as first argument the spinor whose momentum should be set, the second specifies the momentum. It can be one of the following.
\begin{itemize}
\item The four vector in the form of a list {\tt \{E,p1,p2,p3\}}, (note that it is the responsibility of the user to make sure that the vector is indeed an on-shell vector) 
\item two explicit spinors, the first in the $\lambda$ form and the second in the $\tilde\lambda$ form of (\ref{eq:lambda}). This can be used with the {\tt La} and {\tt Lat} functions. If the momentum associated with a spinor is defined using two spinors, these spinors will be used for the numerical evaluation of the spinor products.    
\end{itemize}
If the first argument was not declared as a spinor using {\tt DeclareSpinor}, it will be declared.
 \rule{1cm}{0cm}\vspace{0.2cm}\\
\rule{-1cm}{0cm}\fbox{\parbox{\exboxlength}{
\exampleSp\\
\exampleSp\\
\exampleSp\\
\exampleSp\\
\exampleSp\\
\exampleSp
}
}\vspace{0.1cm}\\
\item[\tt DeclareLVectorMomentum]\myindex{DeclareLVectorMomentum}\rule{1cm}{0cm}\\
One can also define four-vectors that are not associated with a spinor (for example when they are not light-like) with the function {\tt DeclareLVectorMomentum}. It takes two arguments, first the symbol for the vector to be defined and second the value of the four vector in the form of a list.
\vspace{0.2cm}\\
\rule{-1cm}{0cm}\fbox{\parbox{\exboxlength}{
\exampleSp\\
\exampleSp\\
\exampleSp
}
}\vspace{0.1cm}\\
\item[\tt GenMomenta]\myindex{GenMomenta}\rule{1cm}{0cm}\\
The function {\tt GenMomenta[\{s1,...,sn\}]} generates arbitrary on-shell four momenta for the spinors {\tt s1,...,sn}. They are generated so that they sum up to zero momentum. 
\vspace{0.2cm}\\
\rule{-1cm}{0cm}\fbox{\parbox{\exboxlength}{
\exampleSp\\
\exampleSp
}
}\vspace{0.1cm}\\
The function {\tt GenMomenta[\{s1,...,sn\}->\{p0,p1,p2,p3\}]} generates arbitrary on-shell four momenta for the spinors {\tt s1,...,sn}, so that the sum of these momenta is equal to the vector $p=(p_0,p_1,p_2,p_3)$.
\vspace{0.2cm}\\
\rule{-1cm}{0cm}\fbox{\parbox{\exboxlength}{
\exampleSp\\
\exampleSp\\
\exampleSp\\
\exampleSp\\
\exampleSp\\
\exampleSp
}
}\vspace{0.1cm}\\
The momenta can be produced with {\tt GenMomenta} in a reproducible way by seeding the random number generator with the Mathematica command {\tt SeedRandom}.   
\vspace{0.2cm}\\
\rule{-1cm}{0cm}\fbox{\parbox{\exboxlength}{
\exampleSp\\
\exampleSp
}
}\vspace{0.1cm}\\
{\tt GenMomenta} can take as a second argument the precision with which the momenta should be generated. Too large precision slows down the numerical evaluation.
\vspace{0.2cm}\\
\rule{-1cm}{0cm}\fbox{\parbox{\exboxlength}{
\exampleSp\\
\exampleSp\\
\exampleSp
}
}\vspace{0.1cm}\\
\end{description}
%___________________________________________________________________
\subsection{Numerical Functions}
%.........................................................
Once the momenta associated with the spinors are declared or generated, several numerical functions are accessible.
\begin{description}
\item[\tt Num4V]\myindex{Num4V}\rule{1cm}{0cm}\\
The value of the momentum associated to the spinor $a$ is accessible through the function {\tt Num4V}.
\vspace{0.2cm}\\
\rule{-1cm}{0cm}\fbox{\parbox{\exboxlength}{
\exampleSp\\
\exampleSp
}
}\vspace{0.1cm}\\
{\tt Num4V} is linear in the same way as {\tt MP}.
\vspace{0.2cm}\\
\rule{-1cm}{0cm}\fbox{\parbox{\exboxlength}{
\exampleSp\\
\exampleSp
}
}\vspace{0.1cm}
The explicit numerical representation of the spinors are accessible by applying the Mathematica function {\tt N} on the two- or four-dimensional representation, both are represented in a matrix notation. {\tt La}, {\tt CLat}, {\tt USpa} and {\tt USpb} are column vectors, whereas  {\tt CLa}, {\tt Lat}, {\tt UbarSpa} and {\tt UbarSpb} are row vectors.
\vspace{0.2cm}\\ 
\rule{-1cm}{0cm}\fbox{\parbox{\exboxlength}{
\exampleSp\\
\exampleSp\\
\exampleSp\\
\exampleSp\\
\exampleSp\\
\exampleSp\\
\exampleSp\\
\exampleSp\\
\exampleSp\\
\exampleSp\\
\exampleSp
}
}\vspace{0.1cm}\\
Spinor products with or without inserted slashed matrices corresponding to spinor momenta as well as  the invariants {\tt s[a,...,b]} can be evaluated numerically.
\vspace{0.2cm}\\
\rule{-1cm}{0cm}\fbox{\parbox{\exboxlength}{
\exampleSp\\
\exampleSp\\
\exampleSp\\
\exampleSp\\
\exampleSp\\
\exampleSp
}
}\vspace{0.1cm}\\
Once the numerical value of four vectors are defined, one can use them in spinor products with slashed matrices, vector products {\tt MP[p,q]}, {\tt MP2[p]} and invariants {\tt s[p,q]}.
\vspace{0.2cm}\\
\rule{-1cm}{0cm}\fbox{\parbox{\exboxlength}{
\exampleSp\\
\exampleSp\\
\exampleSp\\
\exampleSp\\
\exampleSp\\
\exampleSp\\
\exampleSp\\
\exampleSp\\
\exampleSp\\
\exampleSp\\
\exampleSp
}
}\vspace{0.1cm}\\
\item[\tt PfrmSm2, PfromCSm2]\myindex{PfrmSm2}\myindex{PfrmCSm2}\rule{10cm}{0cm}\\
The four vector can be extracted from the numerical two-dimensional representation of its slashed matrix using the function {\tt PfromSm2} or {\tt PfromCSm2}, depending on whether the matrix is of the {\tt Sm2} or {\tt CSm2} type.
\vspace{0.2cm}\\
\rule{-1cm}{0cm}\fbox{\parbox{\exboxlength}{
\exampleSp\\
\exampleSp\\
\exampleSp\\
\exampleSp\\
\exampleSp
}
}\vspace{0.1cm}\\
\item[\tt PfrmSm4]\myindex{PfrmSm4}\rule{10cm}{0cm}\\ The four vector can be extracted from the numerical four-dimensional representation of its slashed matrix using the function {\tt PfromSm4}.
\vspace{0.2cm}\\
\rule{-1cm}{0cm}\fbox{\parbox{\exboxlength}{
\exampleSp\\
\exampleSp\\
\exampleSp
}
}\vspace{0.1cm}\\

\end{description}

%---------------------------------------------------------------------------
\section{Simple Examples}
%---------------------------------------------------------------------------
In this section we will show three simple examples of the use of the {\tt S@M} package. In the first example, we will re-derive the five gluon MHV amplitude $A^{\mbox{tree}}(1^-,2^-,3^+,4^+,5^+)$ using the BCFW construction \cite{BCFW}, in the second, we compute a box integral coefficient numerically using the quadruple cut technique. The third example illustrates the evaluation of cut integrals using the spinor integration method. The examples of this section are distributed with the package.

\subsection{BCFW Construction}
In this example, we will re-derive the very well-known five-gluon MHV amplitude $A^{\mbox{tree}}(1^-,2^-,3^+,4^+,5^+)$ using the BCFW construction \cite{BCFW}.
The Mathematica notebook file {\tt S@M\_BCFW.nb} containing this example  is distributed with the package.
%%%%%%%%%%%%%%%%%%%%%%%%%%%%%%%%%%%%%%%%%%%%%%%%%%%%%%%%%%%%%%%%%%%%%%
\begin{figure}
%%%%%%%%%%%%%%%%%%%%%%%%%%%%%%%%%%%%%%%%%%%%%%%%%%%%%%%%%%%%
$$
%\myamplifour
\begin{picture}(0,0)(0,0)
\SetScale{0.8}
\SetWidth{0.5}
\Line(0, 0)(-27, -27) \Text(-32,-32)[]{{{\footnotesize $5^+$}}}
\Line(0, 0)(-35, 0)   \Text(-40,0)[]{{{\footnotesize $1^-$}}}
\Line(0, 0)(-27, +27) \Text(-32,+32)[]{{{\footnotesize $2^-$}}}
\Line(0, 0)(+27, +27) \Text(+32,+32)[]{{{\footnotesize $3^+$}}}
\Line(0, 0)(+27, -27) \Text(+32,-32)[]{{{\footnotesize $4^+$}}}
\GOval(0,0)(23,23)(0){.95}
\Text(0, 0)[]{{$A_5$}}
\end{picture}
\hspace*{1.5cm}
=
\hspace*{2.5cm}
\begin{picture}(0,0)(0,0)
\SetScale{0.8}
\SetWidth{0.5}
\Line(0, 0)(-27, -27) \Text(-32,-32)[]{{{\footnotesize $5^+$}}}
\Line(0, 0)(-35, 0)   \Text(-40,0)[]{{{\footnotesize $1^-$}}}
\Line(0, 0)(-27, +27) \Text(-32,+32)[]{{{\footnotesize $|2^-]$}}}
\Line(0, 0)(+35, 0)   \Text(+37,10)[]{{{\footnotesize $Q^+$}}}
\GOval(0,0)(18,18)(0){.95}
\Text(0, 0)[]{{$A_4$}}
\end{picture}
\hspace*{2.0cm}
\times
{1\over s_{34}}
\times
\hspace*{2.3cm}
\begin{picture}(0,0)(0,0)
\SetScale{0.8}
\SetWidth{0.5}
\Line(0, 0)(-35, 0)   \Text(-40,-10)[]{{{\footnotesize $(-Q)^-$}}}
\Line(0, 0)(+27, +27) \Text(+32,+32)[]{{{\footnotesize $|3^+\rangle$}}}
\Line(0, 0)(+27, -27) \Text(+32,-32)[]{{{\footnotesize $4^+$}}}
\GOval(0,0)(17,17)(0){.95}
\Text(0, 0)[]{{$A_3$}}
\end{picture}
$$
\vspace*{0.8cm}
\caption{BCFW construction of the five-gluon MHV amplitude using a $\left|23\right>$ shift.}
\label{fig:bcfw}
%%%%%%%%%%%%%%%%%%%%%%%%%%%%%%%%%%%%%%%%%%%%%%%%%%%%%%%%%%%%
\end{figure}
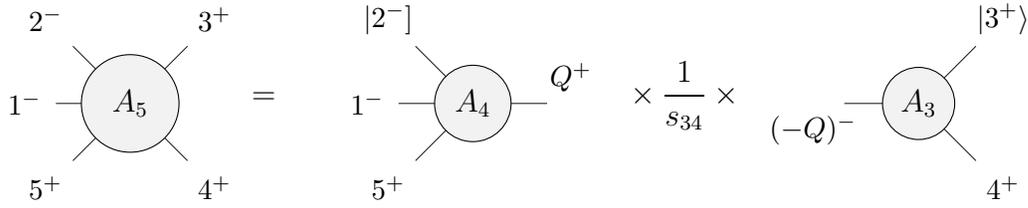
%%%%%%%%%%%%%%%%%%%%%%%%%%%%%%%%%%%%%%%%%%%%%%%%%%%%%%%%%%%%%%%%%%%%%%

We start by defining the MHV amplitudes.
\vspace{0.2cm}\\
\rule{-1cm}{0cm}\fbox{\parbox{\exboxlength}{
\exampleSp\\
\exampleSp
}
}\vspace{0.1cm}\\
Since we know the answer we can use it to check the calculation 
\vspace{0.2cm}\\
\rule{-1cm}{0cm}\fbox{\parbox{\exboxlength}{
\exampleSp\\
\exampleSp
}
}\vspace{0.1cm}\\
We will use the numerical implementation to check the result, so we need to generate a set of five on-shell momenta which sum up to zero momentum.
\vspace{0.2cm}\\
\rule{-1cm}{0cm}\fbox{\parbox{\exboxlength}{
\exampleSp\\
\exampleSp
}
}\vspace{0.1cm}\\
We will use the shift
\[\lambda_2\to\lambda_2,\quad \lambda_3\to\lambda_3 +z \lambda_3,\quad\tilde \lambda_2\to\tilde \lambda_2-z\tilde \lambda_3,\quad\tilde\lambda_3\to\tilde\lambda_3 . \] 
Now we are ready to start the computation. With the shift we chose, there is only one momentum partition contributing namely $\{5,1,2\}\{3,4\}$, see Figure \ref{fig:bcfw}. The amplitude is then given by
\vspace{0.2cm}\\
\rule{-1cm}{0cm}\fbox{\parbox{\exboxlength}{
\exampleSp\\
\exampleSp\\
\exampleSp\\
\exampleSp
}
}\vspace{0.1cm}\\
First we need to find the value of $z$ that puts the propagator 
\[s_{34}(z)=Q(z)^2=(p_3(z)+p_4(z))^2\] 
on-shell.
\vspace{0.2cm}\\
\rule{-1cm}{0cm}\fbox{\parbox{\exboxlength}{
\exampleSp\\
\exampleSp
}
}\vspace{0.1cm}\\
We can check numerically that the formula for the amplitude is right, but for that we need a numerical expression for the spinors $\left|-Q\right]$ and $\left|Q\right>$ associated with the shifted momentum $Q=p_3(z)+p_4(z)$. This can be done with the functions {\tt DeclareSpinorMomentum} and {\tt PformSm2}
\vspace{0.2cm}\\
\rule{-1cm}{0cm}\fbox{\parbox{\exboxlength}{
\exampleSp\\
\exampleSp\\
\exampleSp\\
\exampleSp\\
\exampleSp\\
\exampleSp\\
\exampleSp
}
}\vspace{0.1cm}\\
Now we can numerically evaluate the shifted amplitude and check that we get the right result.
\vspace{0.2cm}\\
\rule{-1cm}{0cm}\fbox{\parbox{\exboxlength}{
\exampleSp\\
\exampleSp\\
\exampleSp\\
\exampleSp\\
\exampleSp\\
\exampleSp
}
}\vspace{0.1cm}\\
Here we see that the numerical result matches with the known MHV result. We can also recover the analytic formula for the MHV amplitude. For that we first need to convert the spinors $\left|-Q\right>$ into $\left|Q\right>$ using 
\[\left|-Q\right>=\pm i\left|Q\right>,\qquad \left|-Q\right]=\pm i\left|Q\right]\]
where the $\pm$ sign convention has no impact here since the spinor $\left|Q\right]$ appears twice. 
\vspace{0.2cm}\\
\rule{-1cm}{0cm}\fbox{\parbox{\exboxlength}{
\exampleSp\\
\exampleSp\\
\exampleSp\\
\exampleSp\\
\exampleSp\\
\exampleSp\\
\exampleSp\\
\exampleSp
}
}\vspace{0.1cm}\\
We can make the result look more familiar using the Schouten identity
\vspace{0.2cm}\\
\rule{-1cm}{0cm}\fbox{\parbox{\exboxlength}{
\exampleSp\\
\exampleSp
}
}\vspace{0.1cm}\\
we can now check that this is the expected known result.
\vspace{0.2cm}\\
\rule{-1cm}{0cm}\fbox{\parbox{\exboxlength}{
\exampleSp\\
\exampleSp
}
}\vspace{0.1cm}\\
%____________________________________________________________________________
\subsection{A Box Coefficient}
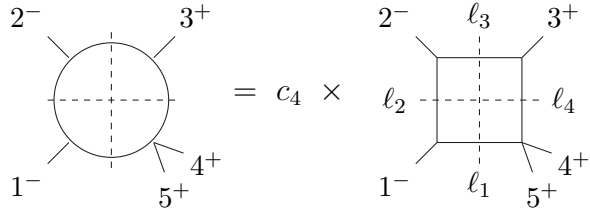
\begin{figure}
%%%%%%%%%%%%%%%%%%%%%%%%%%%%%%%%%%%%%%%%%%%%%%%%%%%%%%%%%%%%
$$
%\myamplifour
\begin{picture}(0,0)(0,0)
\SetScale{0.8}
\SetWidth{0.5}
\Line(-20,-20)(-30,-30) \Text(-32,-32)[]{{{\footnotesize $1^-$}}}
\Line(-20,20)(-30,30) \Text(-32,+32)[]{{{\footnotesize $2^-$}}}
\Line(20,20)(30,30) \Text(+32,+32)[]{{{\footnotesize $3^+$}}}
\Line(20,-20)(34,-25) \Text(+36,-24)[]{{{\footnotesize $4^+$}}}
\Line(20,-20)(25,-34) \Text(+24,-36)[]{{{\footnotesize $5^+$}}}
\Oval(0,0)(27,27)(0)
\DashLine(0,32)(0,-32){3}
\DashLine(-30,0)(30,0){3}
\end{picture}
\hspace*{1.5cm}
=
\ c_4 \ \times \hspace*{1.7cm}
%\mybox
\begin{picture}(0,0)(0,0)
\SetScale{0.8}
\SetWidth{0.5}
\Line(-20,-20)(20,-20)
\Line(20,-20)(20,20)
\Line(20,20)(-20,20)
\Line(-20,20)(-20,-20)
\Line(-20,-20)(-30,-30) \Text(-32,-32)[]{{{\footnotesize $1^-$}}}
\Line(-20,20)(-30,30) \Text(-32,+32)[]{{{\footnotesize $2^-$}}}
\Line(20,20)(30,30) \Text(+32,+32)[]{{{\footnotesize $3^+$}}}
\Line(20,-20)(34,-25) \Text(+36,-24)[]{{{\footnotesize $4^+$}}}
\Line(20,-20)(25,-34) \Text(+24,-36)[]{{{\footnotesize $5^+$}}}
\DashLine(0,30)(0,-30){3}
\DashLine(-28,0)(28,0){3}
\end{picture}
\Text(0,-32)[]{{{\footnotesize $\ell_1$}}}
\Text(-32,0)[]{{{\footnotesize $\ell_2$}}}
\Text(0,+32)[]{{{\footnotesize $\ell_3$}}}
\Text(+32,0)[]{{{\footnotesize $\ell_4$}}}
$$
\vspace*{0.8cm}
\caption{Quadruple-cut for the determination of the coefficient of the box integral $I_4(0,0,0,s_{45})$ in the one-loop five-gluon amplitude, with a gluon propagating in the loop.}
\label{fig:box}
%%%%%%%%%%%%%%%%%%%%%%%%%%%%%%%%%%%%%%%%%%%%%%%%%%%%%%%%%%%%
\end{figure}

We will compute the coefficient of the one mass box $I_4(0,0,0,s_{45})$ of the one-loop five-gluon amplitude $A_5^{\mathrm{1-loop}}(1^-,2^-,3^+,4^+,5^+)$ with a gluon propagating in the loop, see Figure \ref{fig:box}.
The Mathematica notebook file {\tt S@M\_Cut.nb} containing this example  is distributed with the package.
%%%%%%%%%%%%%%%%%%%%%%%%%%%%%%%%%%%%%%%%%%%%%%%%%%%%%%%%%%%%%%%%%%%%%%
%%%%%%%%%%%%%%%%%%%%%%%%%%%%%%%%%%%%%%%%%%%%%%%%%%%%%%%%%%%%%%%%%%%%%%
 
First we generate a momentum configuration with five momenta and define the loop momentum $L$.
\vspace{0.2cm}\\
\rule{-1cm}{0cm}\fbox{\parbox{\exboxlength}{
\exampleSp\\
\exampleSp\\
\exampleSp\\
\exampleSp\\
\exampleSp
}
}\vspace{0.1cm}\\
Note that using the command {\tt SeedRandom[...]} before {\tt GenMomenta} allows us to reproduce the same numerical momentum configuration in another session, so that comparison of numerical results are easier.

Now we define the four propagator momenta {\tt l1,l2,l3,l4} 
\vspace{0.2cm}\\
\rule{-1cm}{0cm}\fbox{\parbox{\exboxlength}{
\exampleSp\\	
\exampleSp\\
\exampleSp\\
\exampleSp\\
\exampleSp
}
}\vspace{0.1cm}\\
and solve for the components of the loop momenta {\tt L} using the on-shell constraints for the four propagators.
\vspace{0.2cm}\\
\rule{-1cm}{0cm}\fbox{\parbox{\exboxlength}{
\exampleSp\\
\exampleSp
}
}\vspace{0.1cm}\\
Using the first solution, we define the spinors corresponding to the propagator momenta (both incoming and outgoing). 
 \vspace{0.2cm}\\
\rule{-1cm}{0cm}\fbox{\parbox{\exboxlength}{
\exampleSp\\
\exampleSp\\
\exampleSp\\
\exampleSp\\
\stepcounter{im}\stepcounter{im}\stepcounter{im}\stepcounter{im}\stepcounter{im}\stepcounter{im}\stepcounter{im}\stepcounter{im}\stepcounter{im}\stepcounter{im}\stepcounter{im}\stepcounter{im}\stepcounter{im}
\dots
}
}\vspace{0.1cm}\\
The output has been shortened for readability. We can now compute the coefficient by inserting these spinors into the amplitudes at each corner of the box. 
 \vspace{0.2cm}\\
\rule{-1cm}{0cm}\fbox{\parbox{\exboxlength}{
\exampleSp\\
\exampleSp\\
\exampleSp\\
\exampleSp\\
}
}\vspace{0.1cm}\\
Where {\tt solh1} and {\tt solh2} are the two contributions corresponding to the two (not obviously vanishing) helicity configurations for the propagator momenta. The first helicity configuration gives a vanishing coefficient. The second solution for the loop momentum yields only vanishing solutions: 
 \vspace{0.2cm}\\
\rule{-1cm}{0cm}\fbox{\parbox{\exboxlength}{ 
\exampleSp\\
\exampleSp\\
\exampleSp\\
\exampleSp\\
\stepcounter{im}\stepcounter{im}\stepcounter{im}\stepcounter{im}\stepcounter{im}
\dots\\ 
\exampleSp\\
\exampleSp\\
\exampleSp\\
\exampleSp\\
}
}\vspace{0.1cm}\\
We can verify that the result is as expected \cite{FiveGluons}
 \vspace{0.2cm}\\
\rule{-1cm}{0cm}\fbox{\parbox{\exboxlength}{
\exampleSp\\
\exampleSp\\
\exampleSp\\
\exampleSp
}
}\vspace{0.1cm}\\
\subsection{Spinor Integration}
\begin{figure}
%%%%%%%%%%%%%%%%%%%%%%%%%%%%%%%%%%%%%%%%%%%%%%%%%%%%%%%%%%%%
$$
%\myamplifour
\begin{picture}(0,0)(0,0)
\SetScale{0.8}
\SetWidth{0.5}
\Line(-20,-20)(-30,-30) \Text(-32,-32)[]{{{\footnotesize $1^-$}}}
\Line(-20,20)(-30,30) \Text(-32,+32)[]{{{\footnotesize $2^+$}}}
\Line(20,20)(30,30) \Text(+32,+32)[]{{{\footnotesize $3^-$}}}
\Line(20,-20)(30,-30) \Text(+32,-32)[]{{{\footnotesize $4^+$}}}
\Oval(0,0)(27,27)(0)
\DashLine(0,32)(0,-32){3}
\end{picture}
\hspace*{1.5cm}
=
\ c_2 \ \times \hspace*{1.7cm}
%\mybox
\begin{picture}(0,0)(0,0)
\SetScale{0.8}
\SetWidth{0.5}
\Line(-20,0)(-30,-10) \Text(-32,-15)[]{{{\footnotesize $1^-$}}}
\Line(-20,0)(-30,10) \Text(-32,+15)[]{{{\footnotesize $2^+$}}}
\Line(20,0)(30,10) \Text(+32,+15)[]{{{\footnotesize $3^-$}}}
\Line(20,0)(30,-10) \Text(+32,-15)[]{{{\footnotesize $4^+$}}}
\Oval(0,0)(20,20)(0)
\DashLine(0,32)(0,-32){3}
\Text(0,-32)[]{{{\footnotesize $\ell_2$}}}
\Text(0,+32)[]{{{\footnotesize $\ell_1$}}}
\end{picture}
$$
\vspace*{0.8cm}
\caption{Double-cut for the determination of the coefficient of 
the bubble integral $I_2(s_{34})$ in the one-loop 4-point gluon amplitude, 
with a gluon propagating in the loop.}
\label{fig:bubble}
%%%%%%%%%%%%%%%%%%%%%%%%%%%%%%%%%%%%%%%%%%%%%%%%%%%%%%%%%%%%
\end{figure}
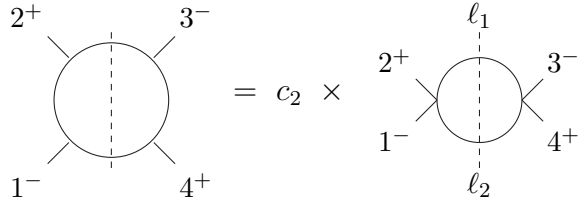

In the following example we discuss how {\tt S@M} can be used to perform
the evaluation of a double-cut by means of the spinor-integration method described in~\cite{BBCF,BFM}. 
The Mathematica notebook file {\tt S@M\_SpinorIntegration.nb} containing this example  is distributed with the package.
In particular, we consider the cut in the $s_{34}$-channel of the one-loop four-gluon amplitude
$A_4^{\rm one-loop}(1^-,2^+,3^-,4^+)$ with a gluon running around the loop.
The $s_{34}$-cut has two contributions, according to the choice of the
internal helicity. We will only take one of the
two possibilities into account, namely 
$$
C_1 = \int d\Phi \  
         A^{\rm tree}(1^-,2^+,\ell_1^-, \ell_2^+) \times
         A^{\rm tree}(\ell_2^-,\ell_1^+,3^-,4^+) \ ,
$$
where $\d\Phi$ is the standard Lorentz invariant two-body phase-space,
see Figure \ref{fig:bubble}
%%%%%%%%%%%%%%%%%%%%%%%%%%%%%%%%%%%%%%%%%%%%%%%%%%%%%%%%%%%%%%%%%%%%%%
%%%%%%%%%%%%%%%%%%%%%%%%%%%%%%%%%%%%%%%%%%%%%%%%%%%%%%%%%%%%%%%%%%%%%%

We begin with the definition of the tree-level amplitude, and of the
cut-integral
\vspace{0.2cm}\\
\rule{-1cm}{0cm}\fbox{\parbox{\exboxlength}{
\exampleSp\\
\exampleSp\\
\exampleSp\\
\exampleSp\\
\exampleSp\\
\exampleSp\\
\exampleSp\\
\exampleSp\\
\exampleSp\\
\exampleSp
}
}\vspace{0.1cm}\\
By using the momentum conservation $\ell_1 - \ell_2 = k_3 + k_4 \equiv P_{34}$,
one can eliminate the dependence on $\ell_2$, 
and write the integrand just in terms of $\ell_1$
\vspace{0.2cm}\\
\rule{-1cm}{0cm}\fbox{\parbox{\exboxlength}{
\exampleSp\\
\exampleSp\\
\exampleSp\\
\exampleSp
}
}\vspace{0.1cm}\\
Then we redefine the loop spinor variables according to the 
CSW~\cite{CSW} prescription: 
\bea
|\ell_1\rangle &=& \sqrt{t} |\lambda\rangle \ , \\
|\ell_1] &=& \sqrt{t} |\lambda] \ , \\
\int d\Phi &=& \int d\lambda \int {t \ dt \over \spab \lambda.P_{34}.\lambda} \ 
\delta\left(t - {s_{34} \over \spab \lambda.P_{34}.\lambda} \right) \\
\int d\lambda &=& \int_{|\lambda] = |\lambda\rangle^*} \dea \deb \ ,
\eea
\vspace{0.2cm}\\
\rule{-1cm}{0cm}\fbox{\parbox{\exboxlength}{
\exampleSp\\
\exampleSp\\
\exampleSp\\
\exampleSp
}
}\vspace{0.1cm}\\
We can perform the $t$-integration by substituting the value of $t$ as imposed
by the $\delta$-function \ ,
\vspace{0.2cm}\\
\rule{-1cm}{0cm}\fbox{\parbox{\exboxlength}{
\exampleSp\\
\exampleSp\\
\exampleSp\\
\exampleSp
}
}\vspace{0.1cm}\\
Now we choose to integrate by parts in $|\lambda]$ and to extract 
residue in $|\lambda\rangle$.
To that aim we have to cast the integrand in a suitable form.
We perform the reduction on the $|\lambda]$-variable by means of the
Schouten identities,
\vspace{0.2cm}\\
\rule{-1cm}{0cm}\fbox{\parbox{\exboxlength}{
\exampleSp\\
\exampleSp\\
\exampleSp\\
\exampleSp
}
}\vspace{0.1cm}\\
The integrand is thus expressed as sum of four terms according to their $\left|\lambda\right]$ dependence.
%three, containing the dependence on $|\lambda]$, in factors like
\[{\spb \bullet.{\lambda}^{n} \over \spab \lambda.P_{34}.{\lambda}^{n+2}} \qquad  n=0,1,2\quad \mbox{or} \quad {1 \over \spab \lambda.P_{34}.{\lambda} \spab \bullet.P_{34}.{\lambda}}\] 
%and one, in factors like
%\[ 1 \over \spab \lambda.P_{34}.{\lambda} \spab \bullet.P_{34}.{\lambda}\] 
\rule{-1cm}{0cm}\fbox{\parbox{\exboxlength}{
\exampleSp\\
\exampleSp\\
\exampleSp\\
\exampleSp\\
\exampleSp\\
\exampleSp\\
\exampleSp\\
\exampleSp
}
}\vspace{0.1cm}\\
For brevity, we discuss only the integration of the term defined as {\tt Cut[1,4]}.
To keep track of the integration we use the explicit definition of $d\lambda$ given
above,
$d\lambda = \dea \deb = {\tt dea \ * \ deb}$ \ .
After simplifying the integrand with trivial spinor identities,
we integrate by parts in $|\lambda]$, using the
identity
\bea
\deb {\spb \eta.\lambda^n \over \spab \lambda.P_{34}.{\lambda}^{n+2} }
&=& {\dedeb \over (n+1)} 
{\spb \eta.\lambda^{n+1} \over 
\spab \lambda.P_{34}.{\lambda}^{n+1} 
\spab \lambda.P_{34}.{\eta} 
}
\eea
for $n=2$ and $|\eta]=|4]$ \ ,
\vspace{0.2cm}\\
\rule{-1cm}{0cm}\fbox{\parbox{\exboxlength}{
\exampleSp\\
\exampleSp\\
\exampleSp\\
\exampleSp\\
\exampleSp\\
\exampleSp\\
\exampleSp\\
\exampleSp
}
}\vspace{0.1cm}\\
The final integration over $|\lambda\rangle$ can be performed by summing over
the residues at the simple poles in $|\lambda\rangle$.
The expression of {\tt Trm4[1]} has two simple poles at 
$|\lambda\rangle = |2\rangle, |4\rangle$.
We notice that the residue at $|\lambda\rangle = |4\rangle$ vanishes, 
due to the term $\spb 4.{\lambda}^3$ in the numerator.
Therefore the result is given just by the residue at $|\lambda\rangle = |2\rangle$,
\vspace{0.2cm}\\
\rule{-1cm}{0cm}\fbox{\parbox{\exboxlength}{
\exampleSp\\
\exampleSp
}
}\vspace{0.1cm}\\
Two comments are in order.
In this simple case, after the $|\lambda]$-integration, the integrand
contained only simple poles in $|\lambda\rangle$.
In general it may well happen that higher poles are present.
Should this be the case, one can apply the function ${\tt ASchouten}$ in order to single out the simple poles beneath the higher ones, and then take the residue 
in $|\lambda\rangle$. \\
The integration of {\tt Cut[1,3]} and {\tt Cut[1,2]} proceeds along the lines
just outlined.
The result, as for {\tt Cut[1,4]}, will be a combination of 
rational functions of spinor products.
Therefore they all contribute 
to the coefficient of the $\ln(s_{34})$ of the whole amplitude.\\
The integration of {\tt Cut[1,1]} is a bit different.
In order to apply the integration by-parts in $|\lambda]$, one has to 
introduce a Feynman parameters. Then, the integration over the spinor
variables can be carried on as outlined above. Finally, 
one performs the integration over the Feynman parameter.
The last parametric integration is responsible for the rising of the logarithmic terms
of the double-cut, whose coefficient can be directly assigned to the 
(poly-)logarithms of the whole amplitude.
%%%%%%%%%%%%%%%%%%%%%%%%%%%%%%%%%%%%%%%%%%%%%%%%%%%%%%%%%
\section{Conclusion and Outlook}
We have presented the Mathematica package {\tt S@M} whose aim is to provide its user with a tool for performing the basic spinor algebra, plus the spinor-shifts
needed for a very efficient analytic evaluation of scattering amplitudes at tree- and loop-level, accompanied by the support of the numerical evaluation at every computational stage.\\
The basic properties of the functions introduced within 
 {\tt S@M} render it a very flexible program which could be further enriched 
with additional routines designed for more specific tasks.

\section*{Acknowledgments}
We wish to thank Zvi Bern, John Conley, Darren Forde, Thomas Gehrmann, Harald Ita, David Kosower, My Phuong Le and Tommer Wizanski for useful comments on the manuscript and for experimenting with un-mature versions of the package. We thank the Galileo Galilei Institute for Theoretical
Physics for the hospitality and the INFN for partial support during the
completion of this work. The work of D. M. was supported by the Swiss National Science Foundation (SNF) under contracts 200020-109162 and PBZH2-117028 and by the US Department of Energy under contract DE-AC02-76SF00515. The work of P.M. was supported by the Marie-Curie-EIF under the contract
MEIF-CT-2006-0214178.
\clearpage
\appendix
\section{Functions Index }\label{sec:index}
The following table lists the functions provided by the package and the page where they are described.\\

%
% latex creates a Spam.idx, process it with
% makeindex -s myindexstyle Spam
% then replace the pageref with /pageref 
%and fix the ordering: s and SMatrix in Spinor.inf
%then insert two times in the Spam.ind file...
%\end{tabular}
%\end{minipage}
%\begin{minipage}{0.3\textwidth}
%\begin{tabular}{ll}
%Command name & page\\\hline
%end add \\{\tt } at the end of the last columm
\begin{minipage}{0.46\textwidth}
\begin{tabular}{ll}
Command  & page\\\hline
\input{Sam.ind}
\end{tabular}
\end{minipage}
\vspace{1cm}\\
A list of all the functions is stored in the variable {\tt \$SpinorsFunctions}.

%\bibliographystyle{h-elsevier}
%\bibliography{all}
\end{document}